\documentclass[preprint,amsmath]{revtex4}
\usepackage{psfrag}
\usepackage[dvips]{graphicx}
\usepackage{textcomp}

\newcommand{\rme}{\mathrm{e}}
\newcommand{\D}{\mathcal{R}}
\newcommand{\K}{\mathcal{K}}
\newcommand{\ud}{\rm{d}}

\begin{document}

\title{Experimental assessment of drag reduction \\ by traveling waves in a turbulent pipe flow}
\author{F. Auteri}
\author{A. Baron}
\author{M. Belan}
\author{G. Campanardi}
\author{M. Quadrio}
\affiliation{Dipartimento di Ingegneria Aerospaziale del Politecnico di Milano
\\ via La Masa 34 - 20156 Milano, Italy}
\date{\today}

\begin{abstract}
We experimentally assess the capabilities of an active, open-loop technique for drag reduction in turbulent wall flows recently introduced by Quadrio {\em et al.} [J. Fluid Mech., {\bf 627}, 161, (2009)]. The technique consists in generating streamwise-modulated waves of spanwise velocity at the wall, that travel in the streamwise direction.

A proof-of-principle experiment has been devised to measure the reduction of turbulent friction in a pipe flow, in which the wall is subdivided into thin slabs that rotate independently in the azimuthal direction. Different speeds of nearby slabs provide, although in a discrete setting, the desired streamwise variation of transverse velocity.

Our experiment confirms the available DNS results, and in particular demonstrates the possibility of achieving large reductions of friction in the turbulent regime. Reductions up to 33\% are obtained for slowly forward-traveling waves; backward-traveling waves invariably yield drag reduction, whereas a substantial drop of drag reduction occurs for waves traveling forward with a phase speed comparable to the convection speed of near-wall turbulent structures.

A Fourier analysis is employed to show that the first harmonics introduced by the discrete spatial waveform that approximates the sinusoidal wave are responsible for significant effects that are indeed observed in the experimental measurements. Practical issues related to the physical implementation of this control scheme and its energetic efficiency are briefly discussed.

\end{abstract}

\maketitle

\section{Introduction}

Controlling wall-bounded turbulent flows to the aim of reducing the wall-shear stress is an important topic in modern fluid mechanics \cite{gadelhak-2000}. Reducing the turbulent friction has significant beneficial effects in a number of technological and industrial applications where turbulent flows interplay with solid surfaces. Potential benefits are significant both performance-wise and from the environmental viewpoint. 

Drag reduction techniques may be classified as active or passive, depending on whether or not an external input of energy is required. To date, passive techniques did not yield large enough a drag reduction to justify their large-scale deployment in industrial applications. In active techniques, on the other hand, the potential benefits need to overcome the energetic cost of the actuation. For closed-loop techniques, where a feedback law between wall measurements and control inputs is designed to achieve the desired goal, this cost is limited. However, feedback techniques for reducing the turbulent friction currently present such challenges, in terms of both  control theory and technological issues for their physical implementation, that their widespread application is not perceived as imminent. In the present paper we do not consider feedback techniques: the interested reader is referred to reviews of recent work \cite{kim-bewley-2007,kasagi-suzuki-fukagata-2009}. 

The focus of the present paper is on open-loop, predetermined techniques, owing to their greater simplicity and ease of implementation. Among such techniques, modifying wall turbulence through large-scale spanwise forcing \cite{karniadakis-choi-2003}, created either by a wall motion or a body force, is emerging as a very attractive approach. The interest into this kind of forcing is increasing thanks to recent technological development in the search of suitable actuators \cite{gouder-morrison-2009,okochi-etal-2009}. 

One effective way of creating the spanwise forcing is to generate {\em spanwise-traveling} waves of body force, aligned in the spanwise direction,  in the very proximity of the wall. Relevant works in this field are several numerical investigations \cite{du-karniadakis-2000}, \cite{du-symeonidis-karniadakis-2002}, \cite{zhao-wu-luo-2004} as well as an experimental study \cite{itoh-etal-2006}. Conceivable implementations to date are limited to body force in conductive fluids, or are supposed to act at the wall. One experimental test of the spanwise-traveling waves concept is available \cite{itoh-etal-2006}, reporting an indirect estimate of up to 7.5\% drag reduction in a turbulent boundary layer by a flexible sheet undergoing a spanwise-traveling wave motion. The drag estimate was based on the growth rate of the boundary layer momentum thickness. 

A new kind of spanwise forcing has been recently introduced by Quadrio, Ricco \& Viotti \cite{quadrio-ricco-viotti-2009} (hereinafter indicated by QRV09). Based on Direct Numerical Simulations (DNS), they observed that waves of spanwise velocity applied at the wall of a turbulent plane channel flow and {\em traveling in the streamwise direction} are capable of altering the natural turbulent friction significantly, while requiring an extremely limited expenditure of energy  The streamwise-traveling waves considered by QRV09 are defined in the planar geometry, and are described by:
\begin{equation}
u_z(x,t) = A \cos \left( \kappa x - \omega t \right),
\label{eq:waves}
\end{equation}
where $u_z$ is the spanwise ($z$) component of the velocity vector at the wall, $x$ is the streamwise coordinate and $t$ is time, $A$ is the wave amplitude, $\kappa$ is the wave number in the streamwise direction and $\omega= 2 \pi / T$ is the oscillation frequency, with $T$ oscillation period. One important parameter of the waves is their streamwise phase speed $c = \omega / \kappa$. Such waves include and generalize the particular cases of the oscillating wall \cite{karniadakis-choi-2003}, where the wall forcing  is spatially uniform:
\[
u_z(t) = A \cos \left( \omega t \right),
\]
and the stationary transverse waves described by Ref. \cite{viotti-quadrio-luchini-2009}, where the forcing presents the appealing characteristic of being stationary:
\[
u_z(x) = A \cos \left( \kappa x \right) .
\]

\begin{figure}
\centering
\includegraphics[width=\columnwidth]{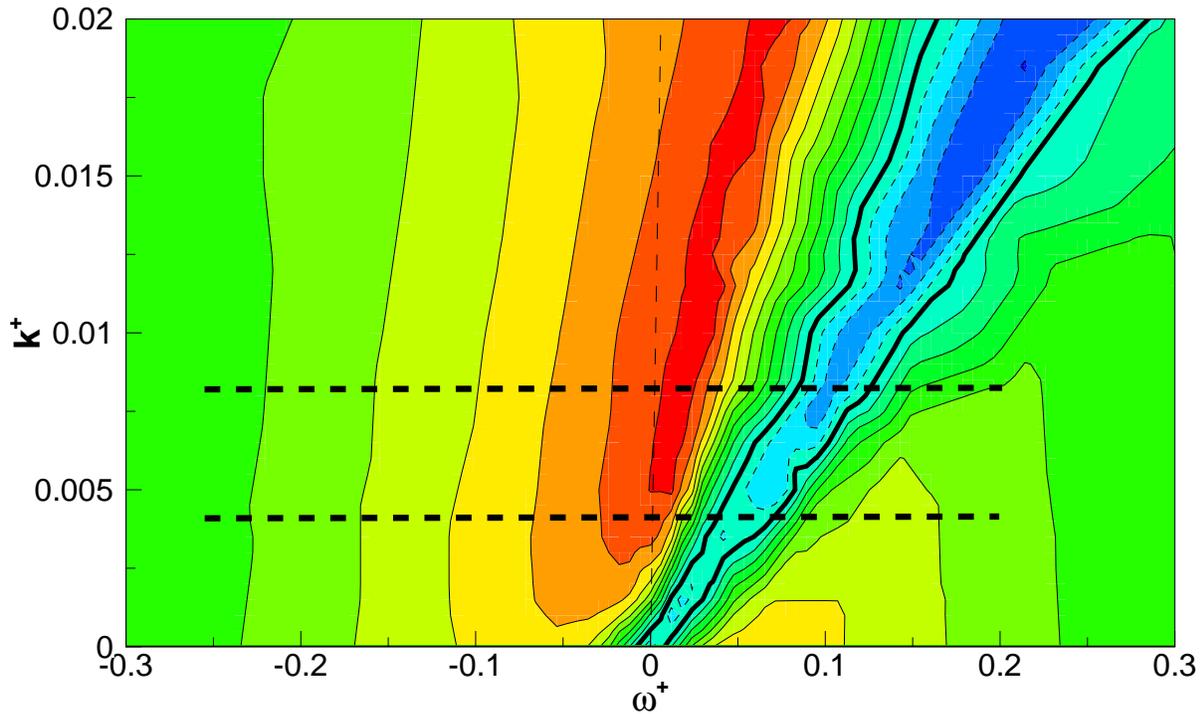}
\caption{Map of friction drag reduction rate $R$ in the $\omega^+ - \kappa^+$ plane, adapted from QRV09 \cite{quadrio-ricco-viotti-2009}. The plane channel flow has $Re_\tau=200$, and the wave amplitude is $A^+=12$. Contours are spaced by 0.05  intervals, zero is indicated by the thick line, and negative values representing drag increase are indicated by dashed lines. The range explored by the present experiment in the circular pipe, described later, is indicated by the two thick dashed horizontal lines.}
\label{fig:DNSresults}
\end{figure}

The effects exerted by the streamwise-traveling waves on the turbulent plane channel flow, as described by QRV09, are summarized in Fig. \ref{fig:DNSresults}, taken and adapted from their paper. The flow has a nominal Reynolds number of $Re_\tau=200$ based on the friction velocity $u_\tau$ and the channel half-width $h$. The amplitude $A^+$ of the waves is kept fixed at $A^+ = 12$ (viscous units, indicated with the $^+$ superscript, are based on $u_\tau$ of the reference flow as the velocity scale and on $\nu / u_\tau$ as the length scale, being $\nu$ the fluid kinematic viscosity). The figure reports the effect of waves in terms of drag reduction rate $R$ as a function of their spatial wavenumber and temporal frequency. Drag reduction rate is defined as the ratio between the reduced pumping power when the waves are on and the reference pumping power; in this case it is given by:
\[
R = 1 - \frac{C_f}{C_{f,0}}
\]
where $C_f$ is the friction coefficient, and the subscript 0 indicates the reference flow without wall forcing. Only half of the $\omega^+ - \kappa^+$ plane is plotted, thanks to the symmetry of the results w.r.t. the origin of the plane. The oscillating wall corresponds to the horizontal axis, whereas the stationary waves correspond to the axis $\omega^+=0$. The effects of the waves change significantly in nature when moving on the $\omega^+ - \kappa^+$ plane. The largest drag reduction (more than 45\% for this value of $A^+$) is observed for slowly forward-traveling waves over a wide range of not-too-large wavelenghts $\lambda^+ = 2 \pi /\kappa^+$. Such waves, at low $Re$, are capable of relaminarizing the flow. At smaller $\kappa^+$, the maximum drag reduction pertains to backward-traveling waves when the wavelength of the waves exceeds $\lambda^+ \approx 1500$. At a well-defined value of the phase speed $c^+$ (indicated in this plot by the inverse of the slope of straight lines passing through the origin), the effect of the waves suddenly becomes that of drag increase. The straight line with $c^+ \approx 11$ identifies the locus of maximum drag increase. The large drag reductions brought about by the waves, combined with the small amount of energy required to obtain them, result in a largely positive overall energy budget. 
In particular, following the notation introduced by Kasagi \cite{kasagi-hasegawa-fukagata-2009}, the waves at $A^+=12$ yield a maximum drag reduction rate of $R=0.48$ and a net saving, computed by taking into account the input power required to support the waves against the viscous resistance of the fluid, of $S=0.18$. The documented best gain $G$, i.e. net energy gain measured in terms of units of input energy, resides at lower $A^+$ and is $G \approx 10$ with $S \approx 0.2$, although is believed that higher values can be found by further exploring the parameter space.
Lastly, this is a large-scale forcing, where the range of scales of best performance is wide, so that the physical wavelength of the waves is, within certain limits, a free design parameter. 

The aim of the present work is to obtain an experimental confirmation of the above results, obtained by QRV09 purely on numerical ground. We have thus built a cylindrical pipe facility, where the traveling waves are generated by independent azimuthal oscillations of thin axial pipe slabs. Drag reductions up to 33\% are measured, and the observed general dependence of the drag changes on the wave parameters follows what has been described in the DNS study. However, one essential difference between numerical simulations and the experiment, namely the discrete spatial waveform used in the experiment to approximate the sinusoidal waves, will be shown to be responsible for features of the experimental data that are absent in the DNS observations.

The paper is organized as follows. In Section \ref{sec:setup} the experimental setup is addressed, by describing its mechanical (\ref{sec:mechanics}) and electronic (\ref{sec:electronics}) components; \ref{sec:parameters} describes the flow parameters and the waveforms that are used in the experiment; experimental procedures and setup validation are described in \ref{sec:procedures}. Section \ref{sec:results} reports the main results, whereas Section \ref{sec:discussion} is devoted to an in-depth discussion; in particular, the issue of using a discrete spatial waveform instead of a sinusoidal one is addressed in \ref{sec:discrete-waveform}, and some practical issue related to the physical implementation on the waves are addressed in \ref{sec:practical-issues}, with emphasis on their energetic efficiency. Lastly, a concluding summary is given in Section \ref{sec:conclusions}.

\section{Experimental setup}
\label{sec:setup}

The geometry of a pipe with circular cross-section is chosen as a testbed to explore the effects of the streamwise-traveling waves experimentally. Although this choice precludes a close comparison with the DNS study by QRV09, that is carried out in a plane channel, the cylindrical geometry has been preferred: in a circular pipe, the waves can be implemented more easily, owing to the naturally periodic spanwise (azimuthal) direction. The traveling waves are thus waves of azimuthal velocity $u_\theta(x,t)$ that travel along the axial $x$ direction. The space-time distribution of $u_\theta$ is expressed by:
\begin{equation}
u_\theta(x,t) = A \cos \left( \kappa x - \omega t \right) .
\label{eq:waves-cyl}
\end{equation}

\begin{figure}
\includegraphics[width=\columnwidth]{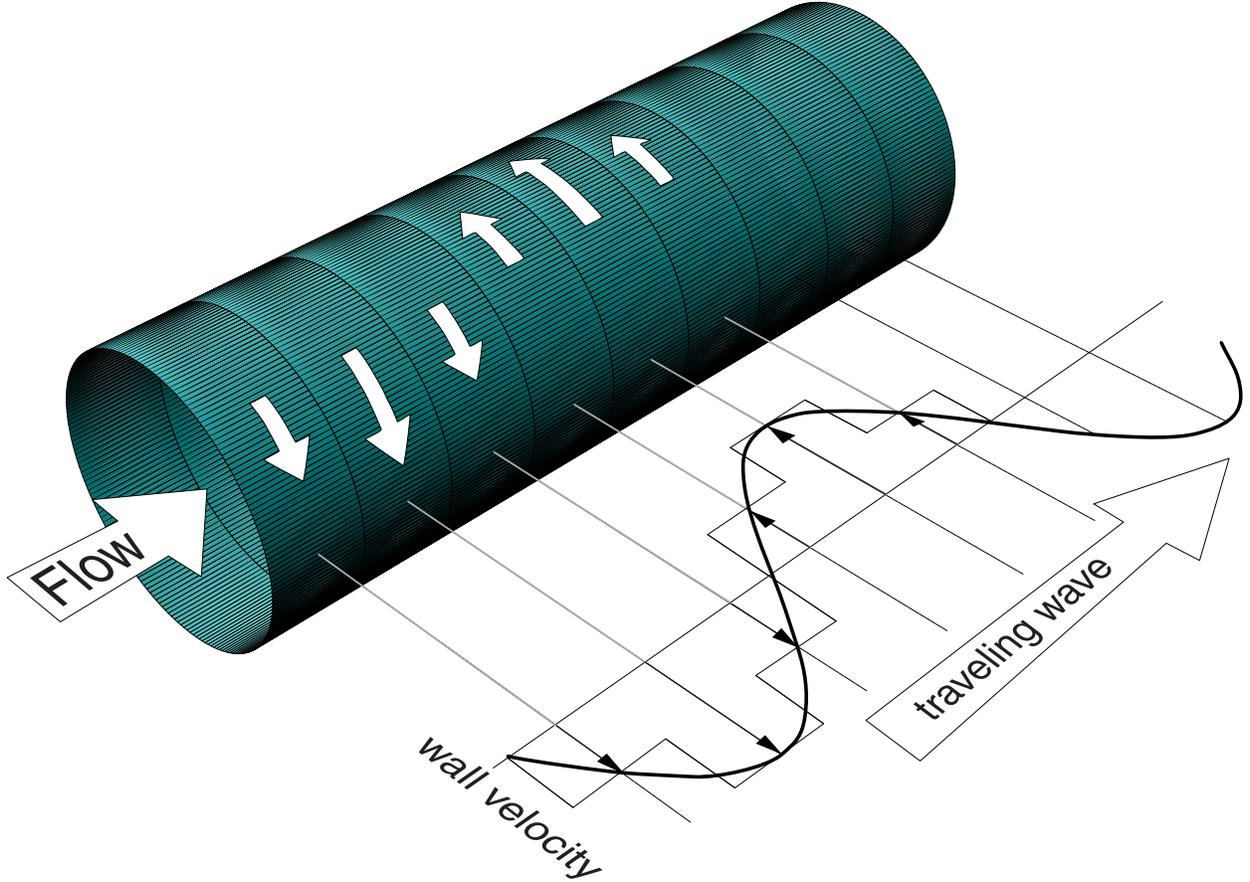}
\caption{Graphical representation of the traveling-wave concept: the desired space-time variation of the transverse wall velocity is achieved through independent alternate motion of adjacent pipe slabs.}
\label{fig:slices} 
\end{figure}

Several ways to implement such a wall forcing can be conceived in a laboratory flow, the simplest example being perhaps a streamwise-modulated transverse blowing through thin streamwise-parallel slots. We instead choose the approach of moving the wall: the spatio-temporal variations involved in Eq. (\ref{eq:waves-cyl}) are thus enforced through a time- and space-varying azimuthal (rotational) speed of the pipe wall. While the sinusoidal dependence on time is easily implementable, the sinusoidal variation along the streamwise direction necessarily requires to be discretized. In a physical setup, this can only be achieved by imposing different rotation rates to different thin longitudinal slabs of the pipe, as illustrated in Fig. \ref{fig:slices}. In the following, our setup will be described in detail.

\subsection{Layout and mechanical setup}
\label{sec:mechanics}

\begin{figure}
\includegraphics[width=\columnwidth]{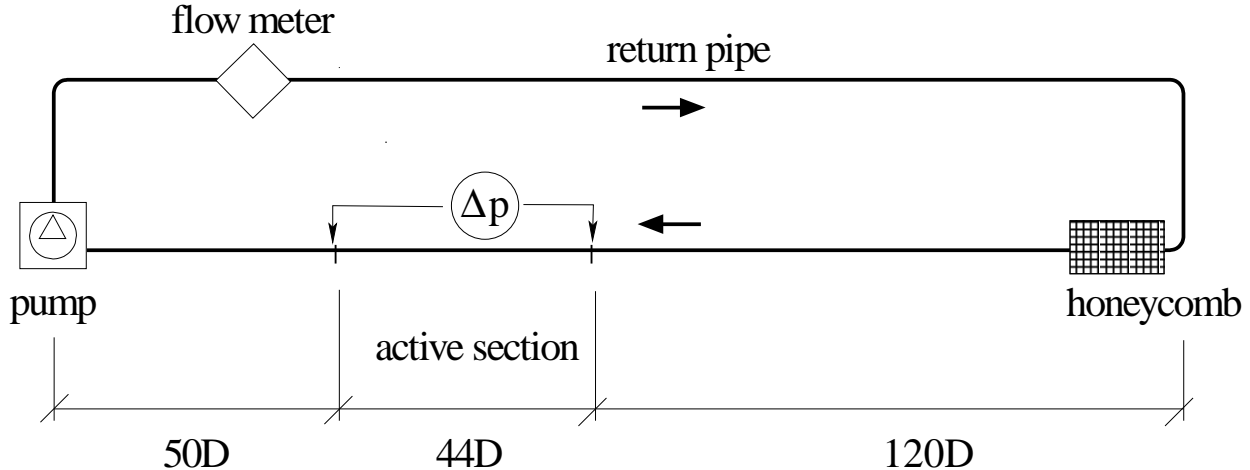}
\caption{Layout of the closed-circuit water pipe facility. The pressure drop $\Delta p$ is measured across the active section, 44$D$ in length, that is preceded by a 120$D$ straight section beginning with honeycomb, and followed by another 50$D$ straight section that connects to the pump. The return pipe hosts the flow-meter device.}
\label{fig:layout} 
\end{figure}

The facility consists of a closed-circuit water pipe, with an inner diameter $D$ of 50 mm $\pm$ 0.05 mm. Its layout is sketched in Fig. \ref{fig:layout}. A straight section 214 $D$ in length includes in its central part the active section where the waves are generated. The active section is preceded by a straight smooth aluminium pipe, 120 $D$ long, to ensure a fully developed turbulent flow. A second smooth straight aluminium section, about 50 $D$ in length, connects the outlet of the active section to the pump and to the return pipe. The facility is given a little slope in order to ease collecting and removing air bubbles from the circuit. A flow-meter is located in the return pipe, to measure the flow rate from which the bulk velocity in the test section is deduced. The mass flow rate is determined by measuring the head loss of an orifice-plate device, designed and built according to the international standard ISO 5167-2 \cite{iso-5167}. Honeycombs are positioned in the circuit well upstream of the active section, in order to avoid swirl. Friction drag is evaluated by measuring the pressure drop between two points located immediately upstream and immediately downstream of the active section. To this purpose, two pressure taps of 0.7 mm diameter, are carefully drilled into the pipe. Details on the employed pressure transducers can be found in the next Section. 
The joints between the various part of the duct are machined with the same accuracy of the moving segments. The rest of the pipe has a tolerance of $\pm 0.15$mm on the inner diameter. The overall misalignement of the moving sections and of the inlet and outlet pipes is less than 0.1\textdegree . The temperature drift after a run of 4 hour is about 0.7\textcelsius. The slope of the facility is about 0.05\textdegree which corresponds to a level difference of about 1 cm between the honeycomb section and the section just upstream of the pump. To remove air from the circuit after filling, when water has reached the laboratory temperature the pump is operated for a few hours. Air bubbles collect in the highest point of the circuit, just upstream of the pump, and are removed through a vent. The air removing procedure is stopped when bubbles are no longer visible in a transparent section of the return circuit.

\begin{figure}
\includegraphics[width=0.6\columnwidth, angle=-90]{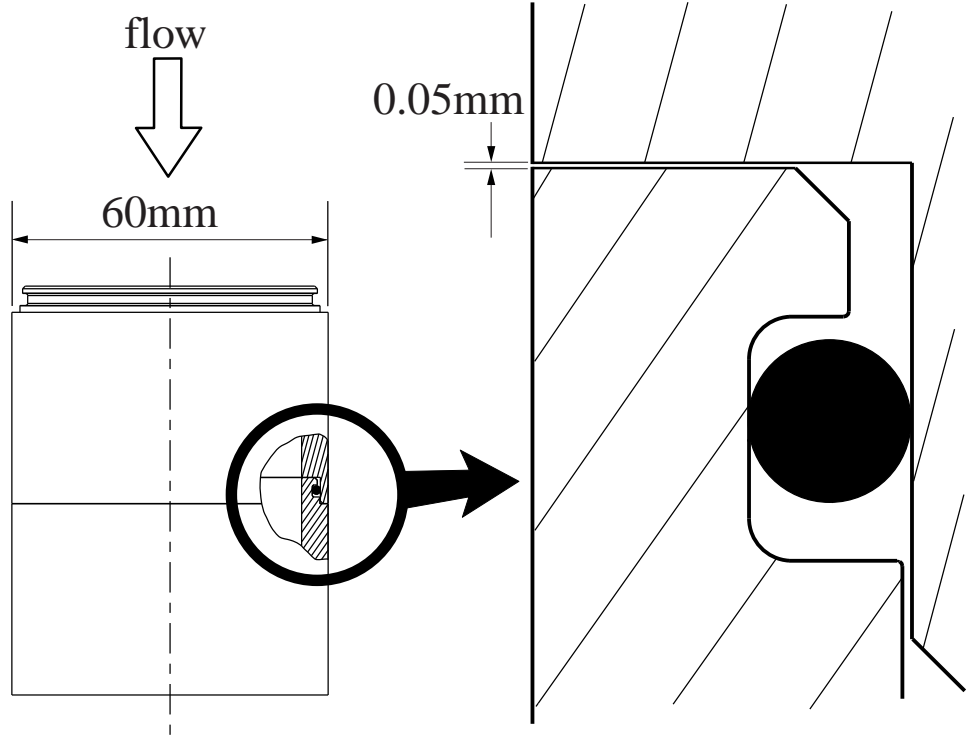}
\caption{Drawing of two adjacent rotating segments and closeup of their interface. The particular shows the sealing O-ring. The 50-micron cavity is filled with water-resistant grease.}
\label{fig:sealing} 
\end{figure}

The longitudinal sinusoidal variation of the transverse velocity in the active section is discretized through up to 6 independently moving segments for each wavelength. The active section is equipped with 60 moving segments: thus, depending on the wave parameters, no less than 10 wavelengths are present along the active section of the pipe. 

Each of the 60 pipe segments is machined from AISI 304 stainless steel, has an axial length of 36.55 mm (about 250 viscous lengths for the typical Reynolds number of the experiment), and is mounted by means of two rolling-contact bearings co-axial to the pipe and aligned to a reference ground steel rail to ensure a highly accurate lineup. Dimensional tolerance and planarity of the rail are within $\pm$ 0.01 mm. Properly sealing the moving segments without applying excessive axial pressure, that would quickly increase mechanical friction above acceptable levels, is obtained thanks to a proper design of the segments, that are fitted with suitable O-rings. A drawing that shows the detail of the interface between two adjacent segments is shown in Fig. \ref{fig:sealing}. After careful mounting, the unavoidable small axial gaps that are present on the inner surface of the pipe at the contact surfaces between moving segments measure about 50--100 microns, that correspond to about one or two thirds of viscous length, or between $0.001D$ and $0.002 D$. The radial discontinuity at the interface is of similar or smaller size. The interfaces are lubricated by water-resistant grease, that serves the multiple purposes of lubricating, filling and sealing the small gaps. 

\begin{figure}
\includegraphics[width=\columnwidth]{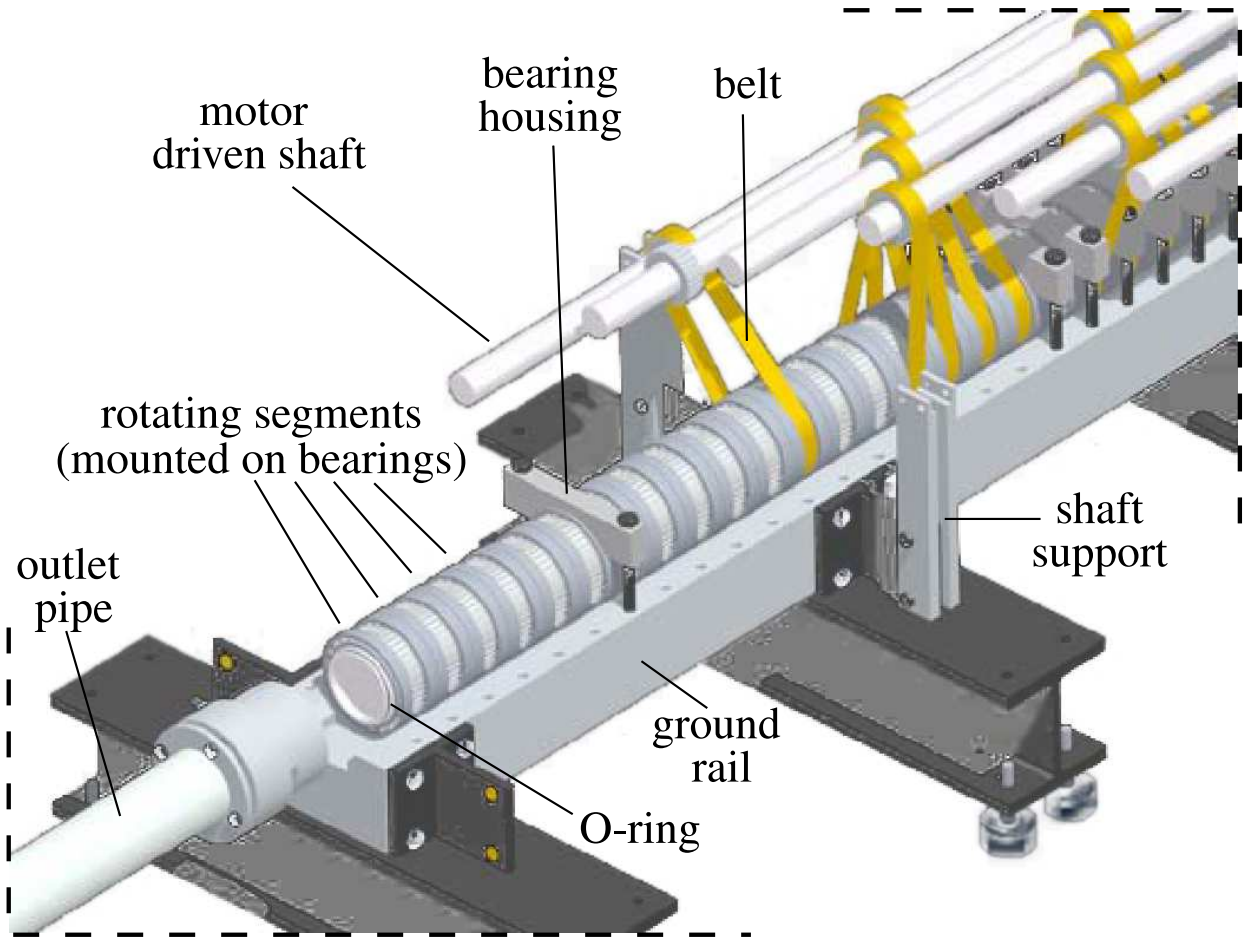}
\caption{Mechanical details of the active portion of the pipe, that is lined on a ground rail. The belts grip on cogwheels and transmit the motion from the motor-driven shafts to the rotating segments. The 6 motors (not shown) are placed at the end of the shafts, 3 at the upstream side and 3 at the downstream side to reduce interference.}
\label{fig:msetup} 
\end{figure}
Fig. \ref{fig:msetup} gives an overall view of the active section and in particular of the transmission system that moves the rotating segments. The 60 segments are driven by 6 independent, nominally identical DC motors 
through shafts and timing belts. If one wavelength has to be discretized with 6 segments, the $i$-th motor drives the segments $(i+6n)$, with $n=0,1,\ldots, 9$. Motors are driven at 24 VDC, are rated 2.7 Nm at 70 W in nominal conditions, and are capable of 27.5 Nm peak at 350 W. The motion is transmitted from the motors to the segments, mounted on rolling bearings, through six shafts. Every shaft is equipped with 10 cogwheels, and moves through timing belts 10 segments rotating with the same phase. 

Due to its particular nature, the setup presents a number of peculiar features. Special care must be taken to prevent water from leaking through the interfaces between the segments, or air and grease from being sucked into the circuit. The solution devised to help avoiding leakages is to control the absolute pressure in the active section, such that it remains just slightly above the external pressure. To this purpose, a vertically-moving reservoir is positioned by a step motor driven by a dedicated controller; the controller receives the instantaneous values of the pressure in the active section and of the external pressure, and acts to modify the vertical position of the reservoir such that the active section pressure remains larger than the external pressure by 500 Pa regardless of the working conditions. 

Galvanic corrosion arises when surfaces of parts made by different metals are in contact in presence of water. Corrosion problems are particularly severe for steel-aluminium contacts. We employ cathodic protection of the entire facility by connecting a suitable power supply to an immersed copper bar acting as the anode and to the metal parts of the circuit as the cathode.

\subsection{Control and instrumentation}
\label{sec:electronics}

\begin{figure}
\includegraphics[width=\columnwidth]{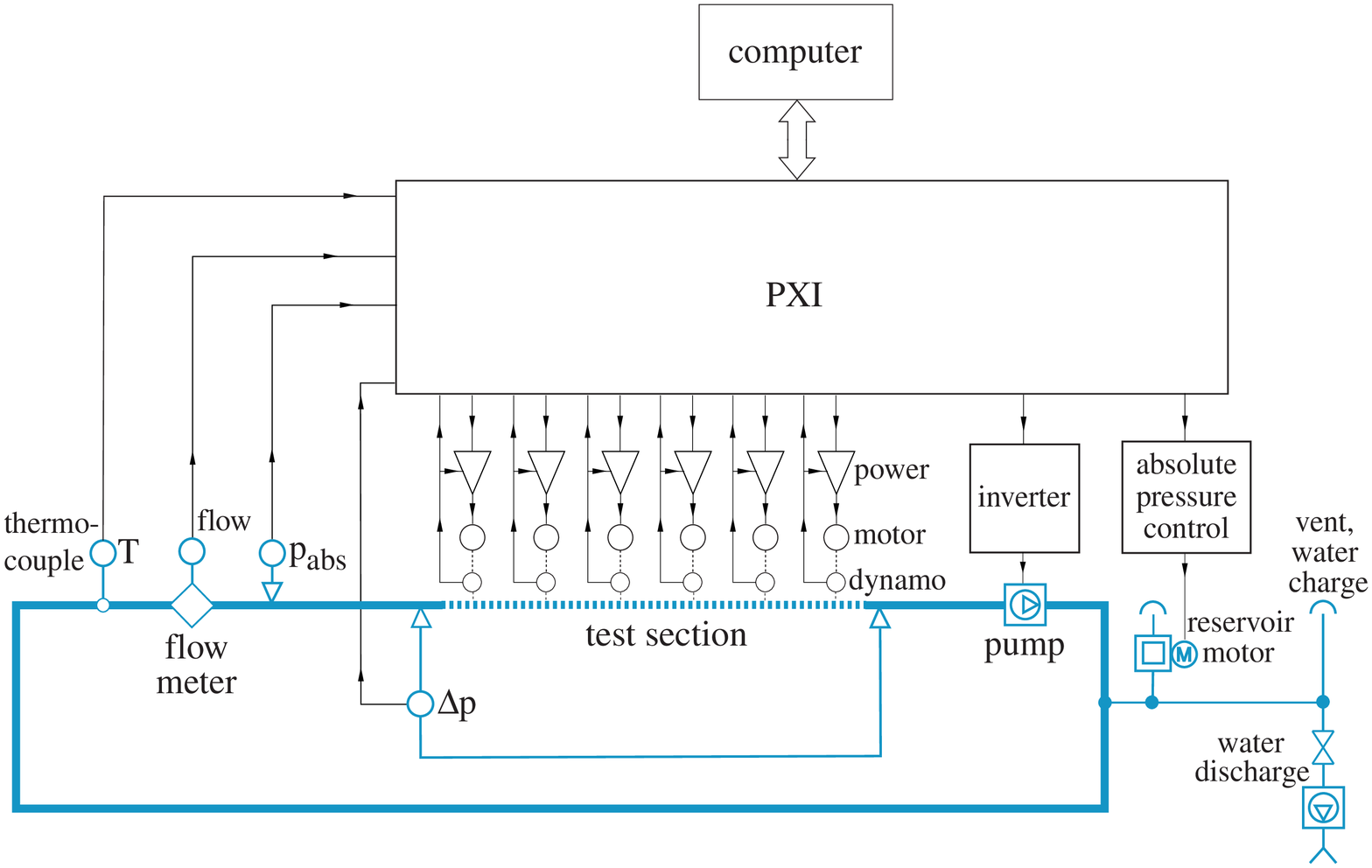}
\caption{Block diagram of the experimental setup. The experiment computer controls both the signal acquisition board and the power stage that drives the segment motors.}
\label{fig:diag} 
\end{figure}

The complete block diagram of the experimental setup is sketched in Fig.\ref{fig:diag}. 
All the operations are managed by a digital interface chassis NI PXI-1010, optically connected to the test control computer. Analog signals are acquired and digitized by 16 bit DAQ modules. Reference signals for the motor power controllers are generated by a 16-bit analog output module, that controls the pump inverter too.

The power controllers that drive the motors are purposely designed and built in house for the present experiment. They are double closed-loop power amplifiers, that reach 100 W output power at $44V_{pp}$. A tachymetric dynamo reads the angular speed of each shaft and provides a signal that is used first in an analog loop inside the relevant controller and, after A/D conversion, in a further digital loop through the PXI system. Each controller is made by a summing/subtracting preamplifier driving a linear AB-class power stage. 

The following pressure transducers are employed in the experiment: 
\begin{itemize}

\item
a $-2\div2\,\textrm{mbar} \pm0.1$\% GE-Druck LPM 9481, labeled '$\Delta p$' in the diagram of Fig. \ref{fig:diag}, is used for the pressure drop measurement across the active section. The transducer is automatically zeroed at the start of each measurement run. 

\item
a $0\div20\,\textrm{mbar} \pm0.1$\% GE-Druck LPM 9481, labeled 'flow' in the diagram, 
is used for the flow-meter head loss measurement. This transducer too is automatically zeroed at the start of each measurement run. The flow meter is calibrated against a traceable standard instrument to improve accuracy. 

\item
a $\pm 5''\,\textrm{WC}$ Setra, labeled '$p_{\rm abs}$', reads the pressure difference between the active section and the laboratory. The output of this transducer is used to keep the pressure in the active section slightly above the external one.

\end{itemize}

Finally, a PT100 thermocouple (accuracy $\pm0.2K$), labeled '$T$' in the diagram and located just upstream of the pump section, is employed to measure water temperature in order to adjust the flow rate to obtain the required value of $Re$. 

Note that the entire circuit is filled with water, including transducers. This avoids the presence of menisca in the measurement line; in such tiny pipes, a meniscus would create strongly curved interfacial surfaces and hence errors in measuring pressure due to surface tension.

\subsection{Experimental parameters}
\label{sec:parameters}

With water as the working fluid, the bulk velocity of the flow is set at $U_b=0.092$ m/s so that the nominal value of the Reynolds number equals $Re=4900$ (based on pipe inner diameter of 50 mm, water viscosity at the measured test temperature and $U_b$). Friction velocity is deduced from the head loss across the active section of the pipe when at rest, and is $u_\tau = 6.7$ mm/s. In terms of friction Reynolds number, or Karman number $R^+$ in the turbulent pipe flow, $Re=4900$ corresponds to $R^+ \approx 175$ and thus it is near to the value of $Re_\tau$ employed in the DNS \cite{quadrio-ricco-viotti-2009} for the planar case, namely $Re_\tau=200$. The value of $R^+$ is large enough for the flow to be fully turbulent, and, at the same time, low enough to yield convenient physical dimensions and operating frequencies of the drag-reducing device. In fact, the oscillating frequencies of interest are of the order of 1 Hz; the pressure drop across the moving section is small but still measurable, of the order of $10$ Pa or 1 mm of water column. 

The dashed lines drawn on the map of DNS results presented in Fig. \ref{fig:DNSresults} serve the purpose of visualizing the parameter range that is accessible to the present experiment. The wall forcing can take the wavelength corresponding to three pipe segments, i.e. $s=3$, or six segments, i.e. $s = 6$. (Indeed, moving the segments at $s=2$ is possible too; this mode, for which discretization effects are largest, will be briefly presented in Section \ref{sec:discussion}. The space-uniform motion where all the segments move in phase is not considered in this paper. The same holds for $s=4$ and $s=5$, that would be possible but at the cost of a time-consuming unmounting of the setup.) Of course wavelength is severely quantized in our setup. The three discrete levels are  $\lambda^+=511$ at $s=2$, $\lambda^+=766$ at $s=3$ and $\lambda^+=1532$ at $s=6$. Tab. \ref{tab:parameters} shows the parameters for the discrete waves that are obtained for $s=2$, $3$ and $6$.
\begin{table}
\begin{tabular}{c|rrrrr}
	& $\lambda$ [mm]&  $\lambda^+$	& $\lambda/R$	& $\kappa^+$	&$\kappa R$ \\ 
\hline 			\hline
$s=2$	&  73.1		& 511		& 2.92		&0.0123		& 2.15	\\
$s=3$	& 109.6		& 766		& 4.38		&0.0082		& 1.43	\\
$s=6$	& 219.3		& 1532 		& 8.77	 	&0.0041		& 0.72	\\
\end{tabular} 
\caption{Discretization parameters for the waves tested in the present experiment. $s$ is the number of segments that rotate independently and discretize one wavelength.}
\label{tab:parameters}
\end{table}

The temporal frequency of the waves can be varied continuously, up to a maximum value that is limited by the mechanical and inertia characteristics of the device, by the transmission system and of course by the motors. The system has been designed so that, at the maximum value of $\kappa$, the available frequency range is such that the drag-increasing regime can be observed. The maximum frequency, shown in Fig.\ref{fig:DNSresults} by the extrema of the dashed lines, is about $\omega^+=0.20-0.25$, and corresponds to 1.8 Hz, a frequency still below the mechanical limits of the actuation system. Similar considerations limit the maximum amplitude $A^+$ of the traveling waves, i.e. the maximum azimuthal speed of the pipe segments during their oscillations. We know from QRV09 (see their figure 6a) that drag reduction is monotonic with $A$, and saturates at large $A^+$. For most of the measurements carried out in the present study, this parameter is thus kept fixed at the value $A^+=13.8$ that is very similar to (indeed, slightly larger than) the DNS value of $A^+=12$. 

\subsection{Procedures and validation}
\label{sec:procedures}

The experimental facility, once filled with deionized water, is left running to degas for a few hours. After complete air bubble removal, the actual tests are started. To minimize the effect of pressure fluctuations and electro-magnetic disturbances, quite long settling and acquisition times have been employed. Before each run, an initial settling period of 120 seconds is allowed, after which the friction factor $f_0$ of the steady reference flow is acquired. Then a series of measurements is started, scanning over the entire frequency range for the fixed value of the number $s$ of segments that discretize one wavelength; each measurement point corresponds to an acquisition of 120 seconds at a frequency of 29 kHz, and before the next measurement point, a further 40 seconds of settling time is adopted. For every experimental point, temperature is measured and the flow rate is adjusted to get the correct $Re_b$ before the pipe slabs are set into motion. The entire series of measurements is fully automated, typically consists in about 40 points and thus requires a few hours of run time. Eventually a further acquisition of the reference friction $f_0$ concludes the test run.

The drag-reducing characteristics of the traveling waves are evaluated by measuring the pressure drop $\Delta p$ across the length $L$ of the active section of the pipe, through the definition
\[
f \equiv \frac{2 \Delta p}{\rho U_b^2 L/D} .
\]

When the active section is at rest, the friction factor $f_0$, obtained in the same way, characterizes the reference flow. Its value, averaged over the six measurements obtained before and after the 3 test cases described in tab. \ref{tab:parameters}, is $f_0 = 0.0405$, with a standard deviation of $\pm 0.0013$, i.e. approximately $\pm 3\%$. This value has been observed to remain constant over a time span of a few weeks.

To compare with, $f_0$ can be estimated by using the Prandtl correlation commonly employed to describe the friction factor in a smooth pipe \cite{pope-2000}, i.e.:
\[
\frac{1}{\sqrt{f}}=2 \log \left( \sqrt{f} Re_b \right) - 0.8 .
\]

The correlation at the nominal value of the Reynolds number $Re_b=4900$ yields the value $f_0=0.0376$. We regard the agreement between the two values of $f_0$ as extremely satisfactory for our setup, since at such a low $Re$, the absolute value of $\Delta p$ (a few Pascal) renders the measurement extremely delicate. The agreement is good enough to assess that possible roughness effects, due to the presence of inter-segment gaps, affect the pressure drop marginally at most. To put this matter in perspective, consider that using the Blasius correlation $f_0 = 4 \cdot 0.079 Re_b^{-0.25}$ to predict $f_0$ yields $f_0 = 0.0378$. The step $k$ between segments is $k/D=0.001$ or $k^+=0.45$, and using Colebrook formula with this roughness height yields $f_0=0.0387$. The distributed roughness is at least one order of magnitude smaller.

\section{Results}
\label{sec:results}

\begin{figure}
\includegraphics[width=0.7\columnwidth, angle=-90]{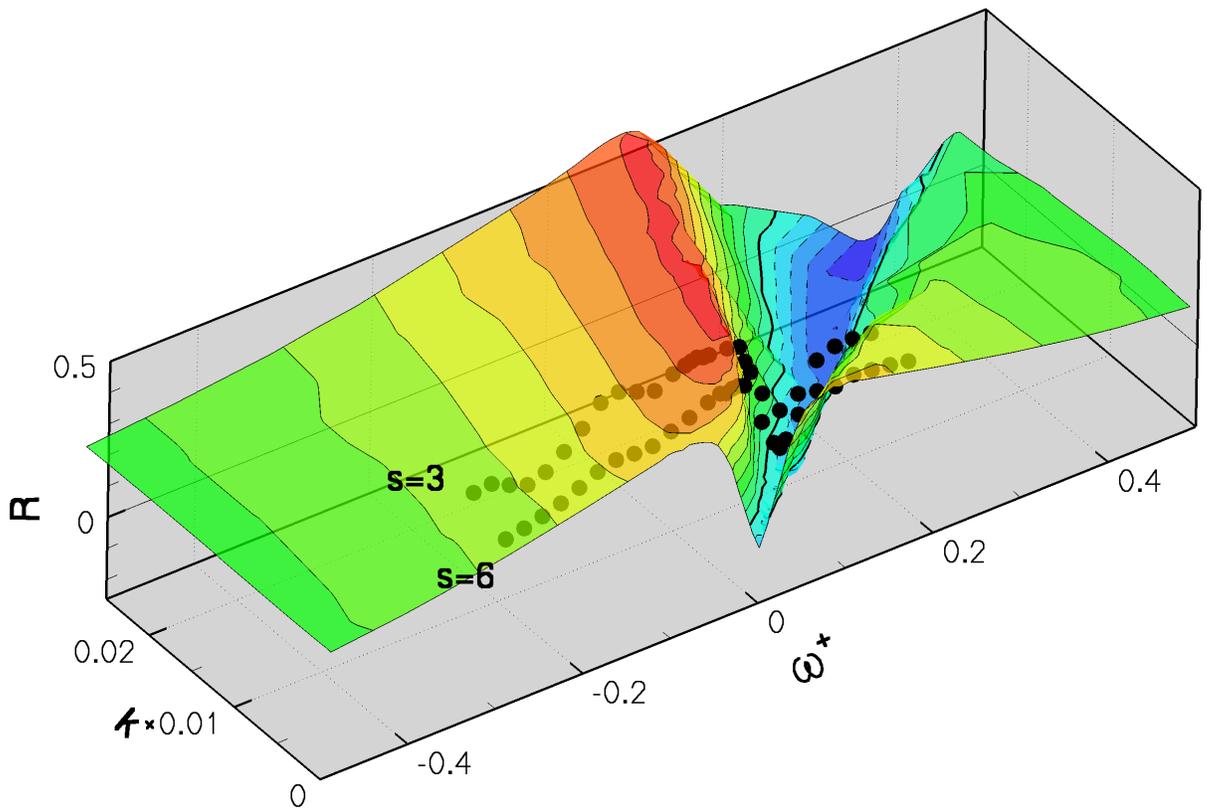}
\caption{Drag reduction rate as a function of wave parameters $\kappa^+$ and $\omega^+$. Comparison between DNS data for planar geometry from QRV09 (colored surface) and present measurements (black dots) for $s=3$ and $s=6$. Contour lines as in Fig. \ref{fig:DNSresults}.}
\label{fig:map-numexp}
\end{figure}

Figure \ref{fig:map-numexp} comprehensively reports the present experimental data, compared with the available DNS data for the plane channel flow. We remind that both the wave amplitude and the value of the Reynolds number of the experiment are comparable, although not identical, to those of the DNS. Changes in friction drag induced by the waves are observed as a function of $\kappa^+$ and $\omega^+$. The DNS dataset taken from QRV09 is represented through the colored, partially transparent surface. The present dataset is shown with the two series of black dots, each corresponding to one frequency sweep of relatively fine step for $s=3$ and $s=6$: the dots thus show how the turbulent friction changes along the horizontal dashed lines in Fig. \ref{fig:DNSresults}. It can be immediately appreciated that our measurements confirm the general dependence of drag reduction on the wave parameters as observed with DNS. The peak of maximum drag reduction, as well as the deep valley where drag reduction drops, are very well caught.

\begin{figure}
\centering
\includegraphics[width=\columnwidth]{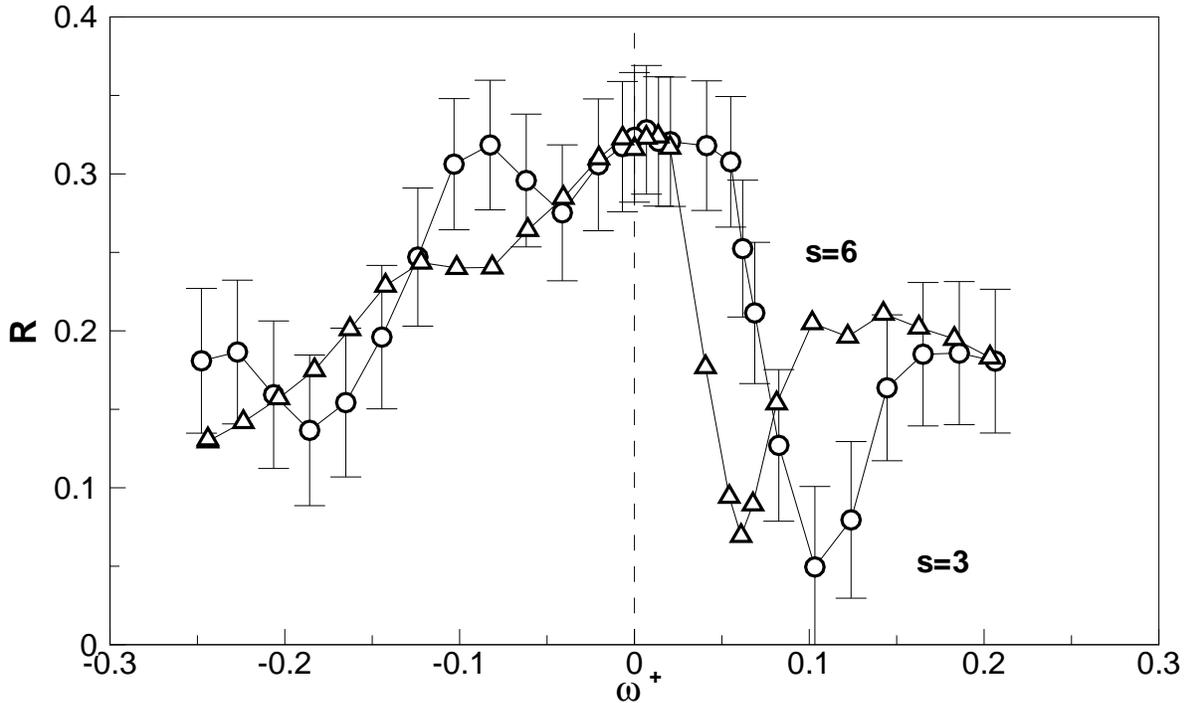}
\caption{Drag reduction rate as a function of the oscillation frequency, for $s=3$ (circles) and $s=6$ (triangles).}
\label{fig:results}
\end{figure}

A more quantitative view at the present experiment's results is offered by Fig. \ref{fig:results}, which plots relative drag reduction data versus $\omega^+$ for the wavelengths corresponding to $s=3$ and $s=6$. The results indeed show the trend that one would expect from DNS. Large reductions of friction drag are successfully measured, whereas an evident drop of drag reduction takes place for a localized range of positive frequencies. The maximum observed reduction of drag is $R \approx 0.33$. In the figure, error bars corresponding to the uncertainties in the measurements of $R$ are reported. They turn out to depend very weakly on the wave parameters, and are of the 
order of $\pm 0.05$. They have been determined by gaussian propagation starting from the accuracy of the pressure transducer measuring across the active section, the accuracy of the temperature probe, which affects density and viscosity, the accuracy of the flow meter pressure transducer, which affects the determination of the bulk velocity, and the geometric tolerances of the pipe and the flow meter itself. We anticipate here that, in addition to such errors, our measurements are affected by another important source of error, notably the spatial transient of drag reduction, that can be considered as a systematic error and will be discussed in \ref{sec:discussion}.

We know from DNS that the frequency at which the maximum drag reduction takes place is weakly wavelength-dependent; the optimal phase speed of the waves goes from a negative one (backward-traveling waves) for very large wavelengths to small and positive for very short wavelengths. Here wavelength have intermediate values, and indeed for $s=6$ the maximum drag reduction is observed for stationary or perhaps slowly-backward traveling waves. When the wavelength is halved at $s=3$ the optimum corresponds to steady or slowly-forward-traveling waves, although it must be said that the level of experimental error makes this maximum comparable to another maximum at negative $\omega^+$. A maximum located at $\omega^+=0.01-0.02$ translates into a phase speed of $c^+ = 1.2-2.4$. 

Both experimental curves present a region, corresponding to positive frequencies and thus to forward-traveling waves, where drag reduction strongly diminishes. As expected from DNS, this local minimum in drag is observed at a nearly constant value of the wave phase speed. The DNS study of QRV09 determined this value to be $c^+ \approx 11$, whereas the present measurements yield $c^+ \approx 12$ for $s=3$ and $c^+=14$ for $s=6$. Moreover, the curve at larger $\kappa^+$ presents larger drag, in agreement with the DNS results.

The general agreement notwithstanding, significant discrepancies do exist between the present measurements and the available DNS dataset, collected for the plane channel. These discrepancies, as well as the reasons that might explain them, will be addressed in the next Section.

\section{Discussion}
\label{sec:discussion}

At a quantitative level, observations from the present experimental campaign do not entirely correspond to the DNS results presented by QRV09. The key differences are: i) the maximum drag reduction observed in the experiment is lower than that reported in the numerical simulations (33\% vs 48\%); ii) although there is a clear indication that drag reduction due to waves drops when their phase speed is near $c^+ \approx 12$, the increase of friction drag above that of the reference flow reported in the numerical simulations is not observed experimentally; iii) the experimental curves (see Fig. \ref{fig:results} and in particular the negative frequency range) present marked wiggles, that are stronger for the case at $s=3$ and that are not observed in the numerical simulations. Apart from the small difference in the flow and forcing parameters (i.e. slightly different values of $Re_\tau$, $A^+$, etc), several arguments are relevant to motivate such discrepancies. 

First of all, the main mechanism driving the modification of turbulent friction drag in plane channel flow, i.e. the presence of a space-time-modulated transverse boundary layer \cite{quadrio-2008, quadrio-etal-2009}, might be qualitatively different from what happens in the cylindrical geometry. Some discrepancies can thus be expected owing to the different geometry. 

Moreover, a laboratory experiment necessarily differs from the idealized setting of a DNS. Most important is the presence of a spatial transient at the beginning of the active section, where the turbulent friction gradually decreases from the unperturbed level of the steady smooth pipe down to the reduced level induced by the wall forcing. Such transient is absent in the indefinite geometry considered in the spatially-periodic DNS. (An analogous recovery transient downstream the end of the active section exists; this is however of no concern for our measurements, since the pressure drop is measured between immediately upstream the beginning and immediately downstream the end of the active section.) This spatial transient has been invoked by Quadrio \& Ricco \cite{quadrio-ricco-2004} to explain the disagreement observed in the early studies on the oscillating wall between DNS results and some underestimated experimental measurements of drag reduction. Only a few experimental measurements of this transient length are available \cite{choi-debisschop-clayton-1998, ricco-wu-2004}, limited to the case of the oscillating wall.

We are unable to measure the length of the transient of the friction over the active section of the pipe, where we do not have physical access for pressure measurement. However, such a transient length has been related to a transient time interval by using the concept of convection velocity of near-wall turbulent fluctuations \cite{kim-hussain-1993, quadrio-luchini-2003}. The DNS by QRV09 contains information on the temporal transient during which the turbulence friction decreases from the initial value to its long-term reduced value: this transient is of the order of hundreds of viscous time units. Since convection velocity is about 10 when expressed in viscous units, the transient length, although dependent on drag reduction parameters, can thus be estimated to be of the order of thousands viscous lengths. Unfortunately, it is difficult to be more quantitative. It is known from previous work \cite{quadrio-ricco-2003} on the oscillating wall that the temporal transient, besides depending on the simulation being run at constant flow rate or constant pressure gradient, presents large damped oscillations before reaching the long-term value; we currently interpret such oscillations, shown to be discretization-independent, as a non-physical effect, reflecting the non-physical periodic boundary conditions employed in a temporal DNS. Indeed, the few experiments that attempted to measure the spatial transient for the case of the oscillating wall \cite{choi-debisschop-clayton-1998, ricco-wu-2004} did not report any oscillation. We can thus safely state that a significant portion of the active section, which is about 15,000 viscous units long, lies in the initial transient region, and that drag reduction measured via pressure drop, that is an integral measure over the entire active length of the pipe, is significantly underestimated. However, we feel unconfortable to go beyond the simple statement that the transient length should extend for $2000-5000$ viscous units. This is inline with the results of a numerical (non-periodic DNS) analysis \cite{fukagata-kasagi-2003} of the very same transient induced by feedback control.

Speaking of transients, it should also be recalled that a spatial transient exists in the drag-increasing regime too. Based on the DNS study by QRV09, its length is shorter than that for the drag reduction, so that this cannot be invoked to explain the present lack of drag increase, although it certainly does contribute to its underestimate.

We suggest that the remaining features of the experimental data, and in particular their wiggles, can be attributed to our setup inevitably yielding a discrete spatial waveform. This effect will be discussed in depth in the following Section. The analysis is of general interest, since any practical implementation of this forcing scheme will necessarily involve discrete actuators. The last Section will discuss practical issues highlighted by the present experiment, and in particular its energetic efficiency.

\subsection{Effects of discrete spatial waveform}
\label{sec:discrete-waveform}

\begin{figure}
\centering
  \includegraphics[width=\columnwidth]{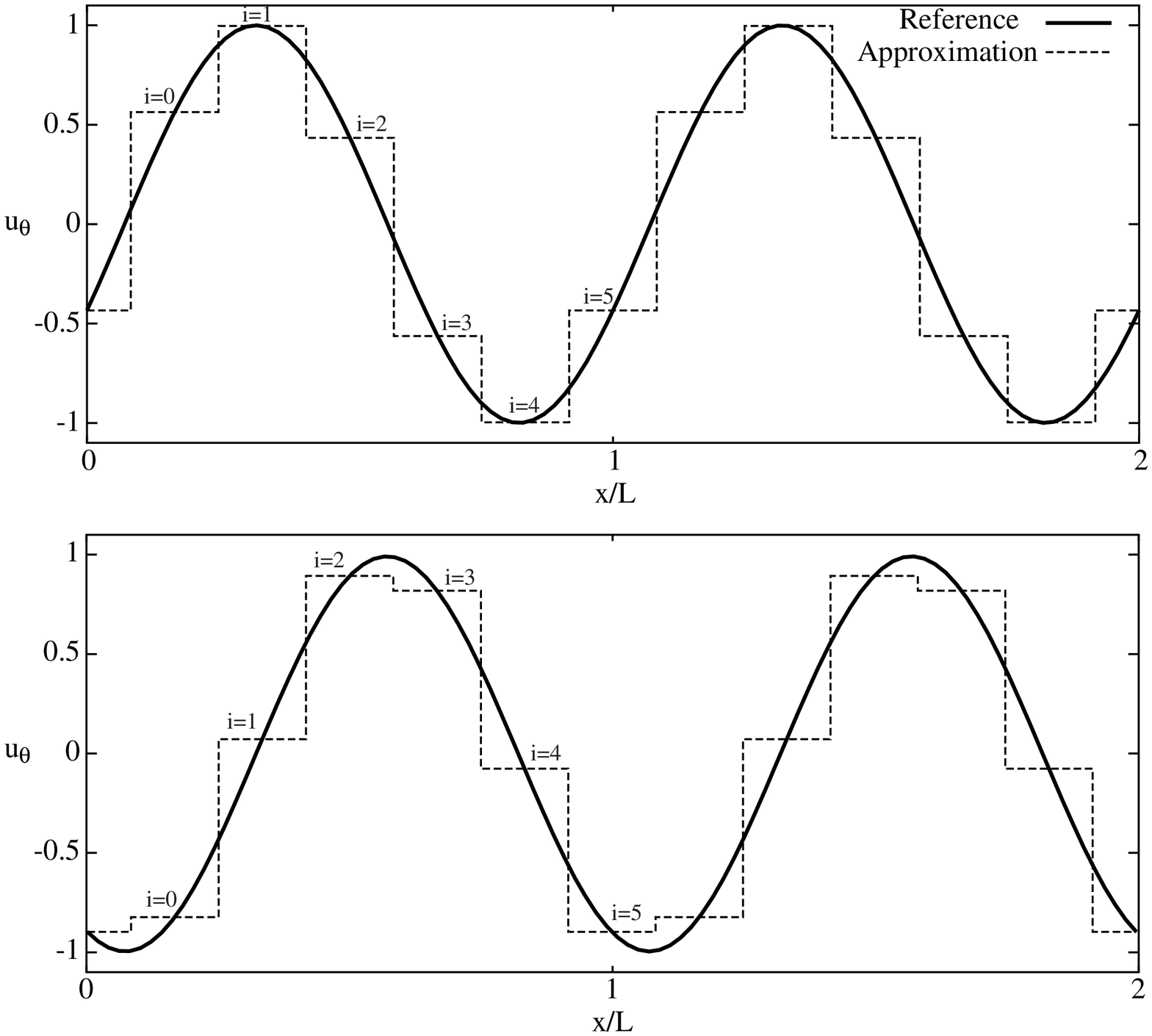}
  \caption{Reference spatial waveform for $u_\theta$ (continuous line) and its discrete approximation $\tilde{u}_\theta$ by 6 segments per period, i.e. $s=6$ (dashed line), at two different times.}
  \label{fig:wavedisc}
\end{figure}

In the physical experiment, the sinusoidal wave (\ref{eq:waves-cyl}) of azimuthal velocity $u_\theta(x,t)$ is approximated by a piecewise-constant periodic function $\tilde{u}_\theta(x,t)$ whose pieces (the pipe slabs) are  fixed in space and change their value (the rotational speed of the slab) with time in a sinusoidal manner, the midpoint of each piece belonging to the continuous sinusoid to be approximated for all time, see Fig. \ref{fig:wavedisc}.

\begin{figure}
\includegraphics[width=\columnwidth]{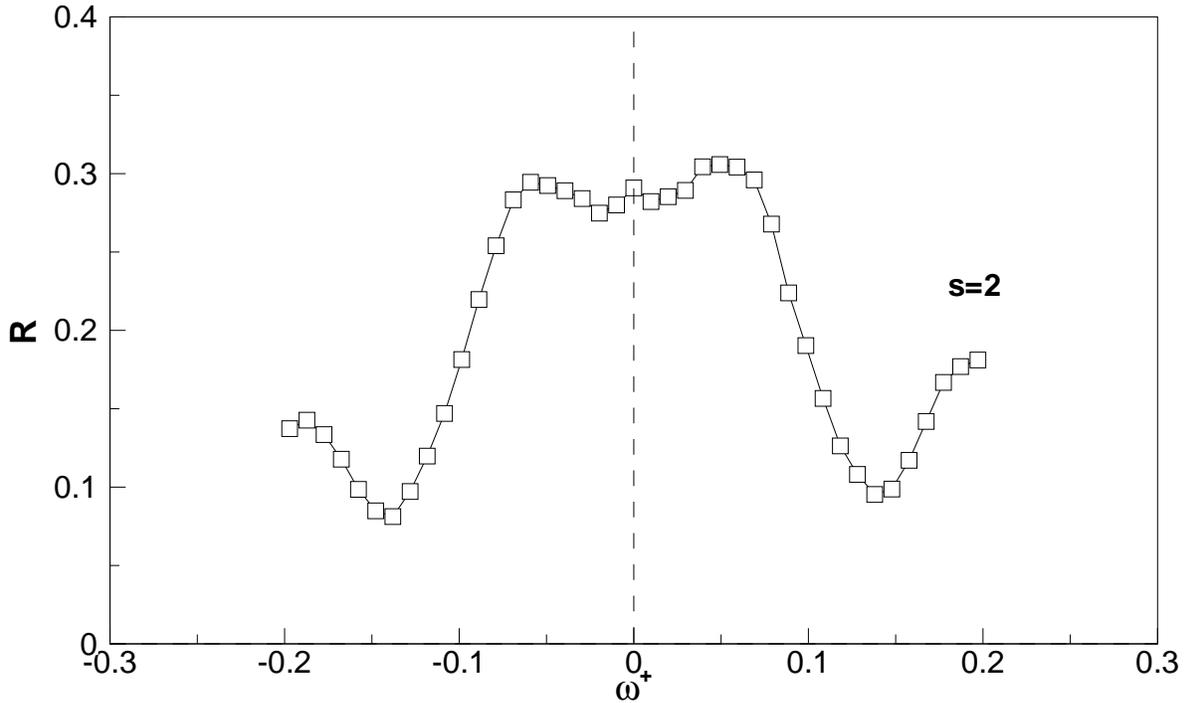}
\caption{Drag reduction rate as a function of the oscillation frequency, for $s=2$.}
\label{fig:p2}
\end{figure}

In the limiting case of $s=2$, i.e. where only two segments are used to approximate the spatial sinusoid, the experimental data vividly highlight the importance of discretization. The drag reduction curve for $s=2$, shown in Fig. \ref{fig:p2}, presents a local maximum at $c^+ \approx 4$ and a local minimum at $c^+ \approx 11$, which is in good agreement with the DNS data from Fig. \ref{fig:DNSresults}. However, the curve is almost perfectly symmetric between positive and negative frequencies, and this may appear as a surprising result. It can be seen clearly from Fig. \ref{fig:DNSresults} that data in the $\omega^+ - \kappa^+$ plane are not symmetric, reflecting an important physical difference between forward- and backward-traveling waves. The symmetry in the measured data is due to the sinusoidal wave being discretized with two segments only. The speed of any two neighboring segments is always in opposite direction, and the wave becomes a standing wave of the type:
\[ 
u_\theta(x,t) = A \cos \left( \kappa x \right) \cos \left( \omega t \right) .
\]
The wave thus has no phase speed, i.e. is the sum of two identical waves traveling in opposite directions. 

Better understanding of the discretization effects is important, since any practical implementation of waves will be discrete and will consequently have to deal with them. To this aim, we carry out a Fourier analysis of the discrete waveform, and start by writing the piecewise-constant periodic function having length $\lambda$ and wave number $\kappa = 2 \pi/\lambda$ as the sum of a number of Heaviside functions equal to the number $s$ of pipe segments per period that are controlled independently from each other ($s=2,3,6$ in our setup where 6 motors and shafts are present). 

Let us denote the $i$-th square function of a total of $s$ by $\tilde{u}_{\theta,i}(x,t;s)$; moreover, let the integer $n$ identify the $n$-th periodic repetition of the discrete waveform, $0 \le n \le (60 / s) -1$. Then, $\forall\,i,\, 0 \leq i < s$, we have
\begin{eqnarray}
&&   \tilde{u}_{\theta,i}(x,t;s) = \\ \nonumber 
&&   \begin{cases}
                              A \sin \left( \omega t - \frac{2 \pi i}{s} \right), &
			      \frac{i-1/2}{s}\, L + n L\leq x < \frac{i+1/2}{s}\,
			      L + nL, \\
			      0, & \textrm{otherwise}.
			   \end{cases}
\end{eqnarray}

First, we neglect border effects, which is equivalent to assuming an infinite number of periodic waveforms, i.e. $n \rightarrow \infty$. The actual waveform can be then assembled as a Fourier expansion of its elementary components, and rearranged by means of trigonometric identities, to yield:
\begin{equation}
\label{eq:expansion}
  \begin{aligned}
&   \tilde{u}_\theta(x,t;s) = A \sum_{m=0}^{\infty}\bigg\{
                          \frac{\sin[(m s + 1)\frac \pi s]}{(m s+1)\frac \pi s}
			   \sin\big[\omega t - \kappa (m s+1)\, x\big] \\[3mm]
&			 + \frac{\sin[((m+1)s-1)\frac \pi s]}{((m+1)s-1)\frac \pi s} 
			   \sin\big[\omega t + \kappa ((m+1)s-1)\, x\big]
			   \bigg\}.
  \end{aligned}
\end{equation}

Regardless of its phase speed, the discrete waveform, once written in the form (\ref{eq:expansion}), appears as composed by the sum of two families of sinusoidal waves, traveling in opposite directions. Each family is a weighted sum of an infinite number of harmonics, their amplitude decreasing with frequency. Focusing on the case $s=2$, the first forward- and backward-traveling harmonics produce at first order the standing wave:
\begin{equation}
\label{eq:fp2}
   \tilde{u}_\theta(x,t;2) \cong \frac{2}{\pi} A \left\{
                           \sin \left( \omega t - \kappa x \right) 
			 + \sin \left( \omega t + \kappa x \right)
			   \right\}.
\end{equation}

The same harmonics for the case $s=3$ yield the composite wave:
\begin{equation}
\label{eq:fp3}
   \tilde{u}_\theta(x,t;3) \cong \frac{3 \sqrt{3}}{2\pi} A \left\{
                           \sin \left( \omega t - \kappa x \right) 
			 + \frac 1 2 \sin \left( \omega t + 2 \kappa x \right)
			   \right\} ,
\end{equation}
whilst the same harmonics for $s=6$ give
\begin{equation}
\label{eq:fp6}
   \tilde{u}_\theta(x,t;6) \cong \frac{3}{\pi} A \left\{
                           \sin \left( \omega t - \kappa x \right) 
			 + \frac 1 5 \sin \left( \omega t + 5 \kappa x \right)
			   \right\}.
\end{equation}

In other words, formula (\ref{eq:fp3}) shows that the idealized sinusoidal wave of amplitude $A$, when discretized with $s=3$, corresponds at leading order ($m$=0) to such a wave with amplitude $3 \sqrt{3} A / 2 \pi \approx 0.83 A$ summed to a second wave, traveling in the opposite direction, with halved wavelength, and amplitude $3 \sqrt{3} A / 4 \pi \approx 0.41 A$. Discretization effects are less severe for the case with $s=6$, where formula (\ref{eq:fp6}) shows that the fundamental harmonic has an amplitude of $3A / \pi \approx 0.95 A$ and the first harmonic is an opposite-traveling wave of 5 times shorter wavelength and amplitude give by $3A/5\pi \approx 0.19A$. Note how the amplitudes of the first harmonics are rather large; indeed, even the next contribution from $m=1$ are remarkably far from zero. Thus, significant discretization effects are to be expected. 

Thanks to the Fourier expansion (\ref{eq:expansion}), the general shape of the experimental curves described in the previous sections can now be discussed to explain, at least partially, the existing differences with DNS data. To this aim, we preliminarily reconsider the surface $\D(\omega,\kappa)$ plotted in Figs. \ref{fig:DNSresults} and \ref{fig:map-numexp} and illustrating the changes in drag reduction rate $R$ introduced by the waves as a function of $\omega$ and $\kappa$. Since $\D(\omega,\kappa)$ is rather smooth, we express it through an integral representation that involves a suitable kernel $\K$ and the generating wave. In the end, this will allow us to give the drag reduction induced by a particular wave an analytical expression, that contains empirical information in the expression for $\K$. 

If the monochromatic traveling wave (\ref{eq:waves-cyl}) of frequency $\omega$ and wavenumber $\kappa$ is considered as the real part of the elementary wave
\begin{equation}
f_{\omega,\kappa}(t,x) = A \Re \left[ \rme^{j \left( \omega t - \kappa x \right)} \right]
\label{eq:monochromatic-wave}
\end{equation}
where $j$ is the imaginary unit, then our {\em ansatz} for $\D(\omega,\kappa)$ is:
\begin{equation}
\D(\omega,\kappa) =
\int\!\!\int \K(\tau,\xi) f_{\omega,\kappa}(\tau,\xi) \, \ud \tau \ud \xi .
\label{eq:spectral-superposition}
\end{equation}

This integral formula tells us that $\D(\omega,\kappa)$ can be obtained by multiplying the (yet to be determined) kernel $\K(\tau,\xi)$ with the traveling wave $f_{\omega,\kappa}$ that forces the flow. We use Eq. (\ref{eq:spectral-superposition}) to empirically determine an analytical expression for $\K$ from the available information on $\D(\omega,\kappa)$, i.e. the drag reduction data obtained by QRV09 through DNS. If a fit is obtained by superposing two (or more) generalized Gaussian functions, then $\K$ is writable as their superposition, owing to the properties of Fourier transforms. We choose to describe $\K$ as the superposition of three generalized Gaussian:
\begin{eqnarray*}
\K(\tau,\xi) = \sum_{j=1}^3 a_0^j \exp \left( a_1^j \tau^2 + a_2^j \tau + a_3^j \tau \xi + a_4^j \xi + a_5^j \xi^2 \right) .
\end{eqnarray*}

\begin{figure}
\includegraphics[width=\columnwidth]{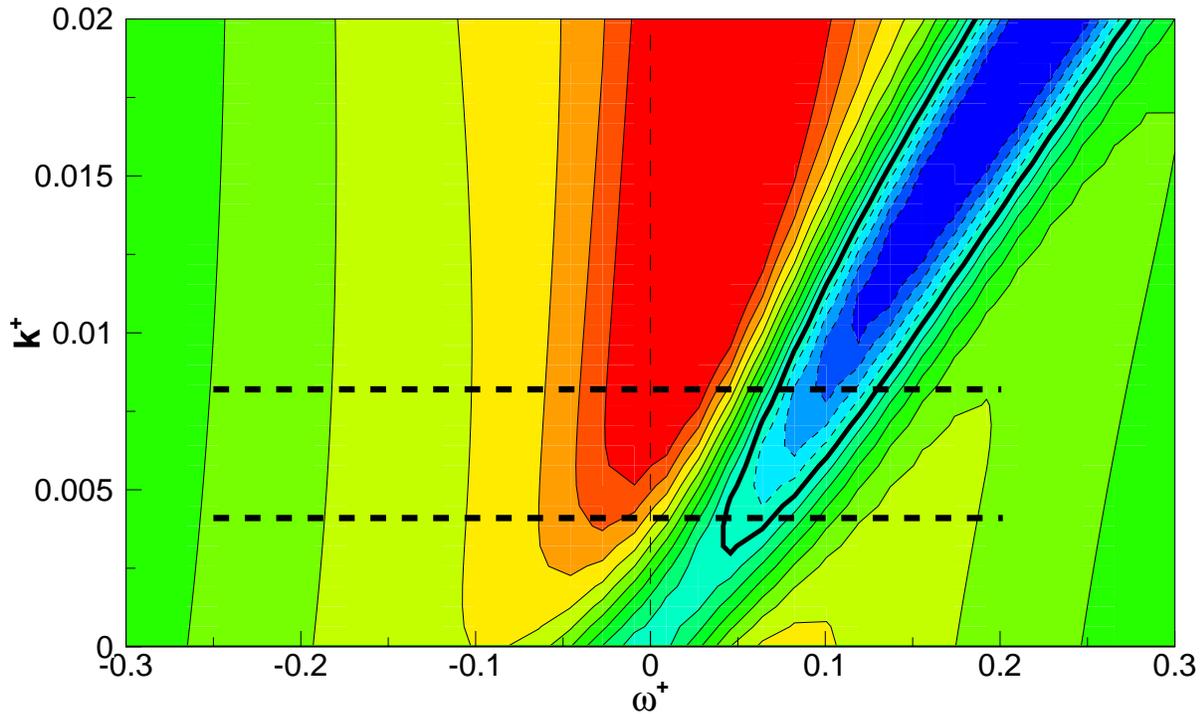}
\caption{Map of the drag reduction rate $\D(\omega,\kappa)$ as reconstructed from the monochromatic traveling wave (\ref{eq:monochromatic-wave}) and the empirically-determined kernel $\K$ defined by the spectral superposition (\ref{eq:spectral-superposition}). Contours as in Fig. \ref{fig:DNSresults}.}
\label{fig:kwmap-0}
\end{figure}

\begin{figure}
\includegraphics[width=\columnwidth]{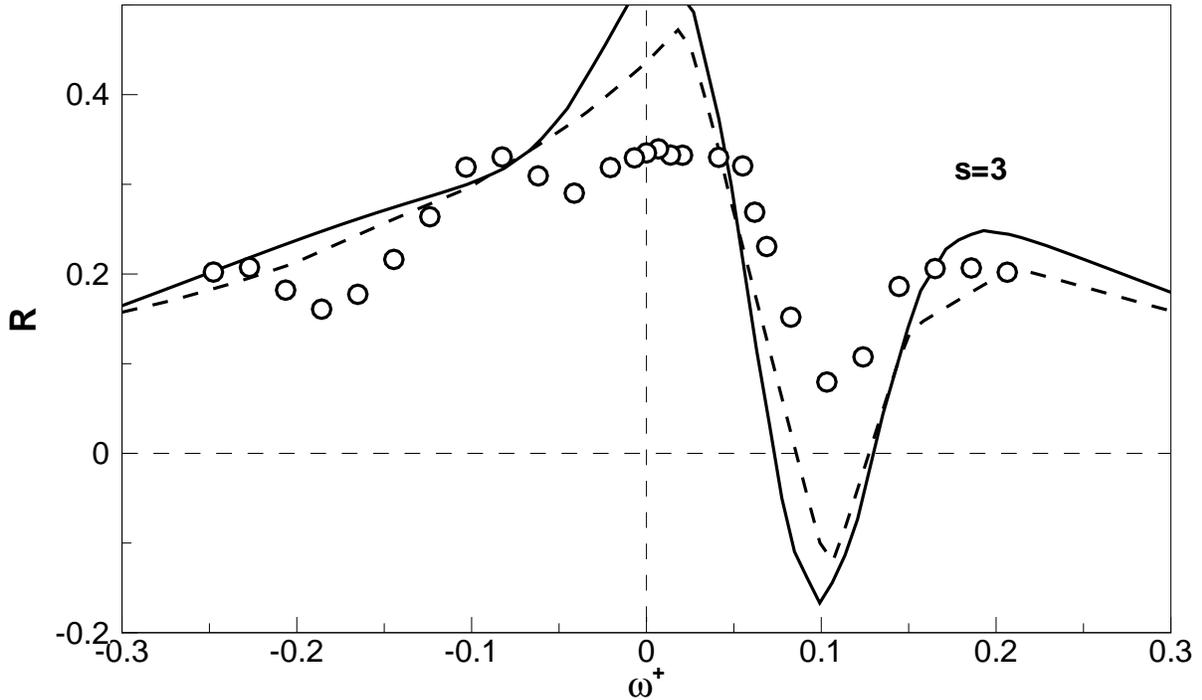}
\caption{Comparison of drag reduction rate at $s=3$ from experiments (symbols), DNS data (dashed line), and fitted convolution of the relevant monochromatic wave (continuous line).}
\label{fig:p3-0}
\end{figure}

\begin{figure}
\includegraphics[width=\columnwidth]{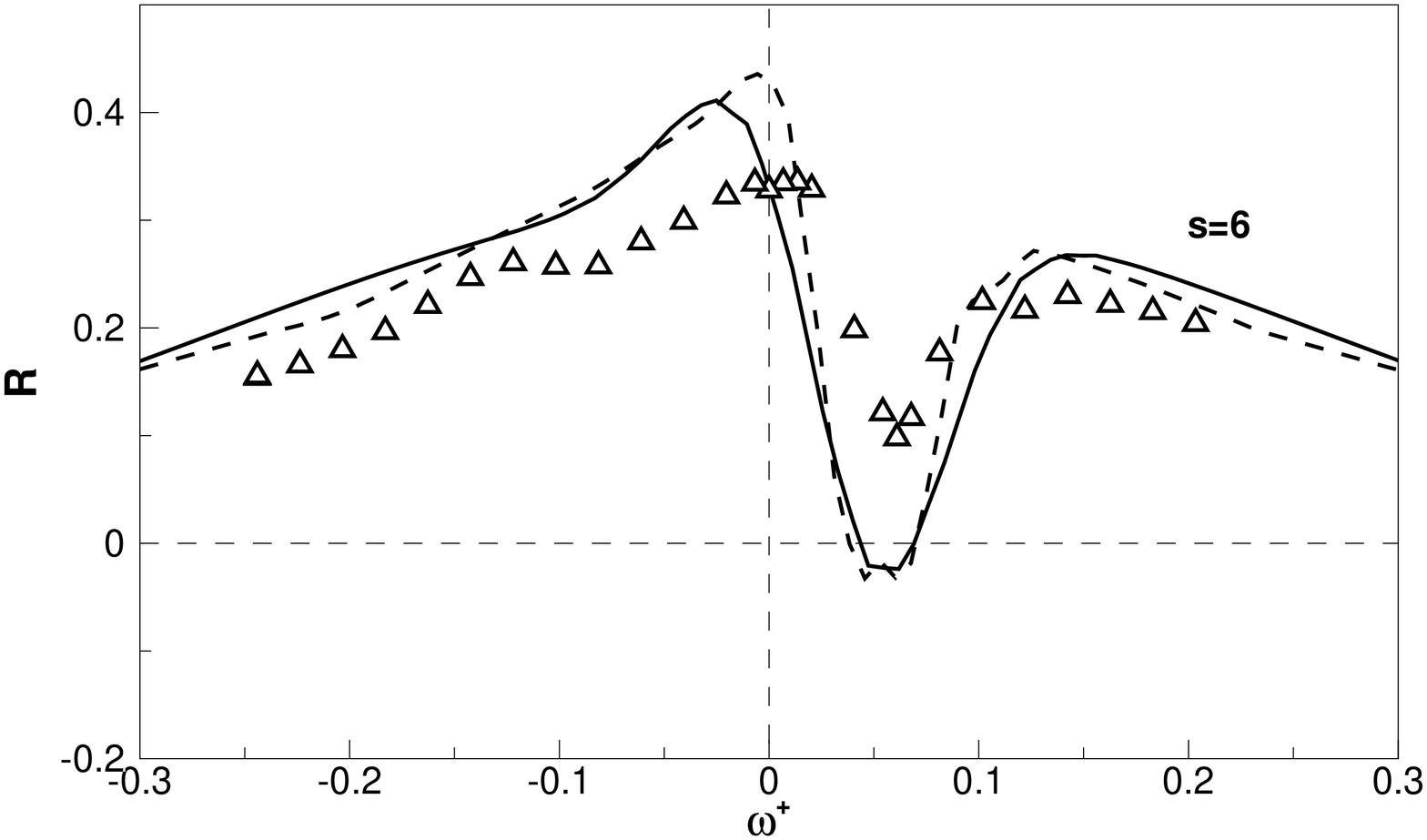}
\caption{Comparison of drag reduction rate at $s=6$ from experiments (symbols), DNS data (dashed line), and fitted convolution of the relevant monochromatic wave (continuous line).}
\label{fig:p6-0}
\end{figure}

A least-squares indirect fit of $\D(\omega,\kappa)$ is used to compute the values of the 18 free complex parameters $a_0^j, a_1^j, \ldots a_5^j, j=1,2,3$ and thus to give the kernel $\K$ an analytical form. Figure \ref{fig:kwmap-0} plots the function $\D(\omega,\kappa)$ as obtained by (\ref{eq:spectral-superposition}), where the analytical form for $\K$ is used. As expected, the general appearance of the map is very similar to the actual map presented earlier in Fig. \ref{fig:DNSresults}, with the largest local error of the fit quite small and confined near the $\omega=0$ region, where the slope of the experimental curve is higher, and confirms that the expression determined for $\K$ represents the DNS data sufficiently well. From a more quantitative viewpoint, Figs. \ref{fig:p3-0} and \ref{fig:p6-0} plot the frequency dependence of drag reduction at the particular wavelengths tested in the experiments ($s=3$ and $s=6$, respectively), and confirm that the fit-based drag reduction curves follow fairly well the observations by DNS, while the present experimental data show some quantitative disagreement as discussed above.

The powerful feature of the model developed so far is that, once an expression for $\K$ is determined, the superposition of different waves can be easily accounted for, since (\ref{eq:spectral-superposition}) does not require the generating wave to be monochromatic. This is equivalent to assuming that the drag-reducing effects of the waves are additive with respect to different waves, i.e. drag reduction rate can be expressed as:
\[
R = 1 - \frac{C_f}{C_{f,0}} = 1 - \frac{u'_w(\omega,\kappa)}{u'_w(0,0)} .
\]
where $u'_w$ is the slope of the velocity profile evaluated at the wall.

\begin{figure}
\includegraphics[width=\columnwidth]{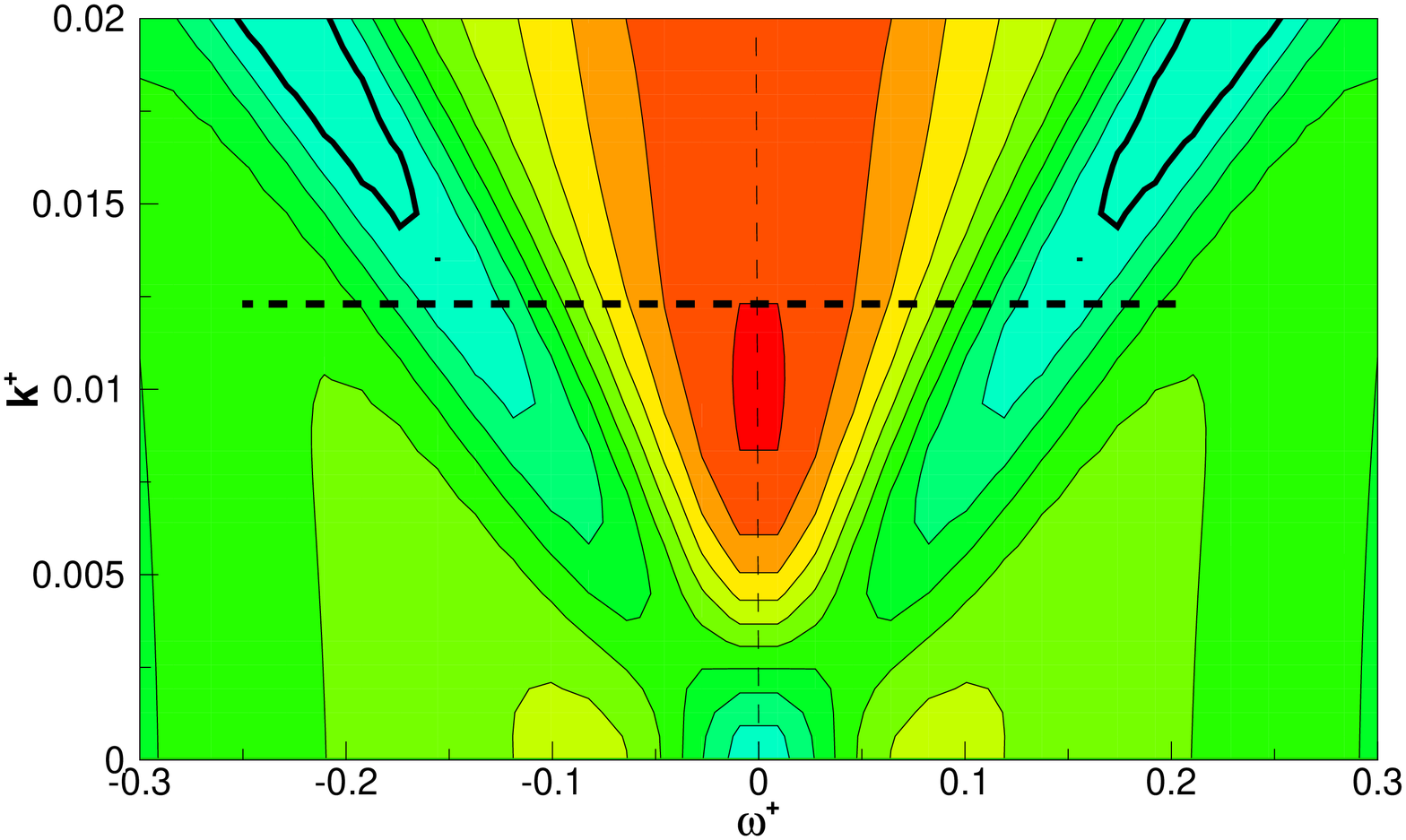}
\caption{Map of the drag reduction rate $\tilde{\D}_\ell(\omega,\kappa; 2)$ obtained by formula (\ref{eq:D-2-1}), i.e. by adding the linear effect of the first harmonic, Eq. (\ref{eq:fp2}). The map presents a left-right symmetry.}
\label{fig:kwmap-p2l}
\end{figure}

In particular, we are interested in investigating the effect of the first harmonics present in Eqns. (\ref{eq:fp2}), (\ref{eq:fp3}) and (\ref{eq:fp6}). For the case $s=2$, where discretization effects are expected to be stronger, use of (\ref{eq:spectral-superposition}) and (\ref{eq:fp2}) yields:

\begin{equation}
\label{eq:D-2-1}
\tilde{\D}_\ell(\omega,\kappa;2) = C_2 \int\!\!\int \K(\tau,\xi) \left[ f_{\omega,\kappa} + f_{\omega,-\kappa} \right] \ud \tau \ud \xi 
\end{equation}
where the tilde indicates that the drag reduction surface is obtained from the discrete wave, the subscript $\ell$ indicates a linear superposition of the effects of the generating waves (\ref{eq:fp2}), and common numerical constants have been factorized into the coefficient $C_2$. The new drag reduction map, shown in Fig. \ref{fig:kwmap-p2l} is symmetric between the right and left half-planes, in agreement with the experimental data reported in Fig. \ref{fig:p2}.

When the case $s=3$ is considered, use of (\ref{eq:spectral-superposition}) and (\ref{eq:fp3}) yields:

\begin{equation}
\label{eq:D-3-1}
\tilde{\D}_\ell(\omega,\kappa;3) = C_3 \int\!\!\int \K(\tau,\xi) \left[ f_{\omega,\kappa} +\frac 1 2 f_{\omega,-2\kappa} \right] \ud \tau \ud \xi .
\end{equation}

\begin{figure}
\includegraphics[width=\columnwidth]{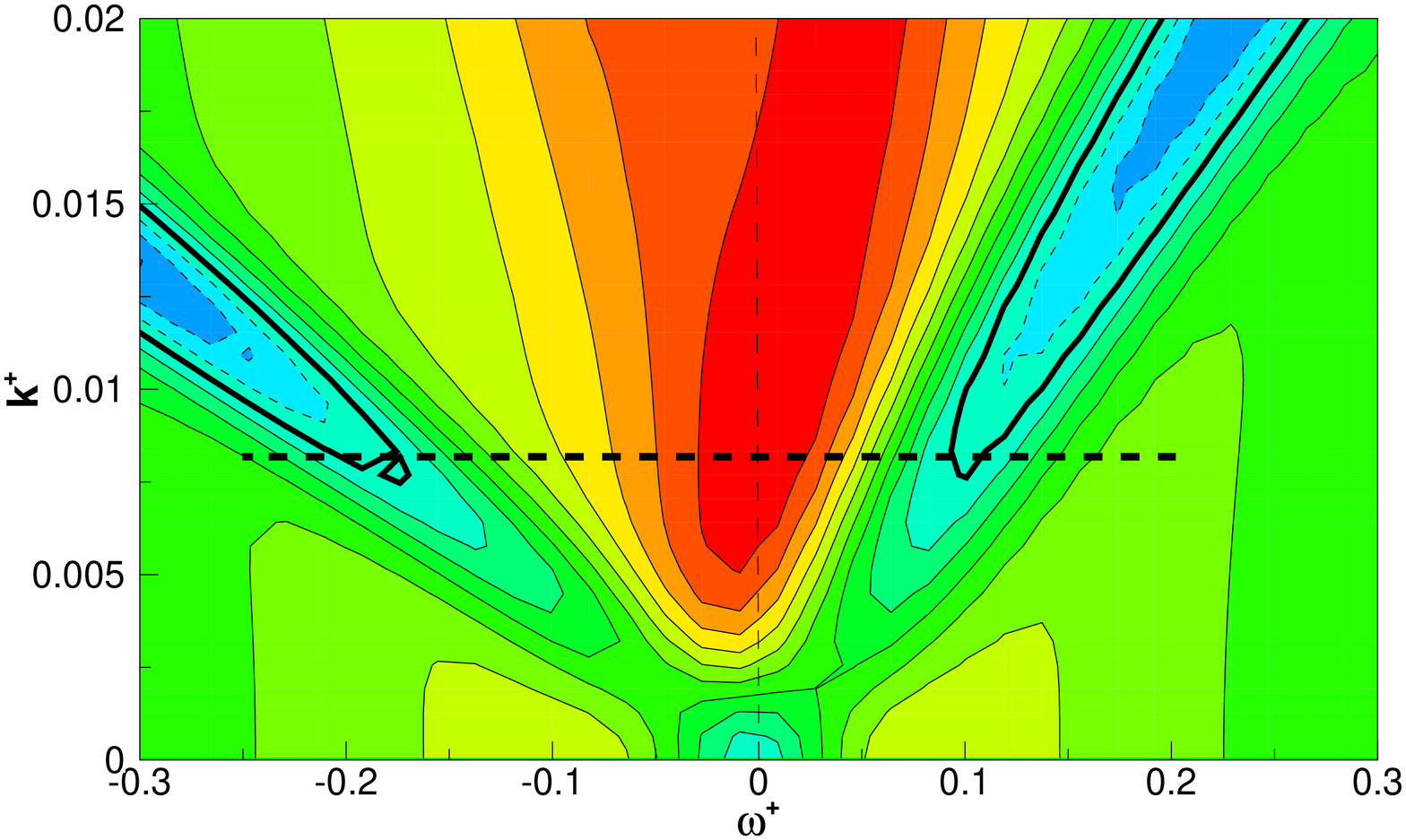}
\caption{Map of the drag reduction rate $\tilde{\D}_\ell(\omega,\kappa; 3)$ obtained by formula (\ref{eq:D-3-1}), i.e. by adding the linear effect of the first harmonic, Eq. (\ref{eq:fp3}).}
\label{fig:kwmap-p3l}
\end{figure}

\begin{figure}
\includegraphics[width=\columnwidth]{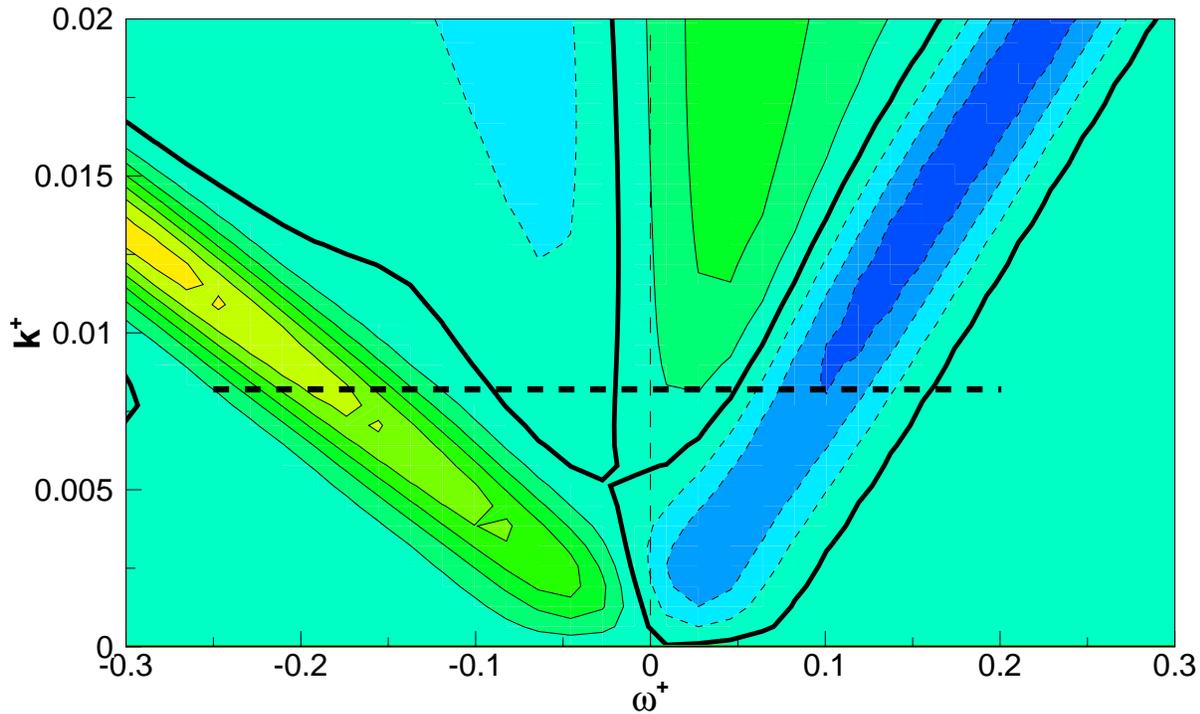}
\caption{Map of the difference $\D(\omega,\kappa) - \tilde{\D}_\ell(\omega,\kappa;3)$ in the $\omega-\kappa$ plane. Contours as in figure \ref{fig:DNSresults}.}
\label{fig:wkdiff}
\end{figure}

\begin{figure}
\includegraphics[width=\columnwidth]{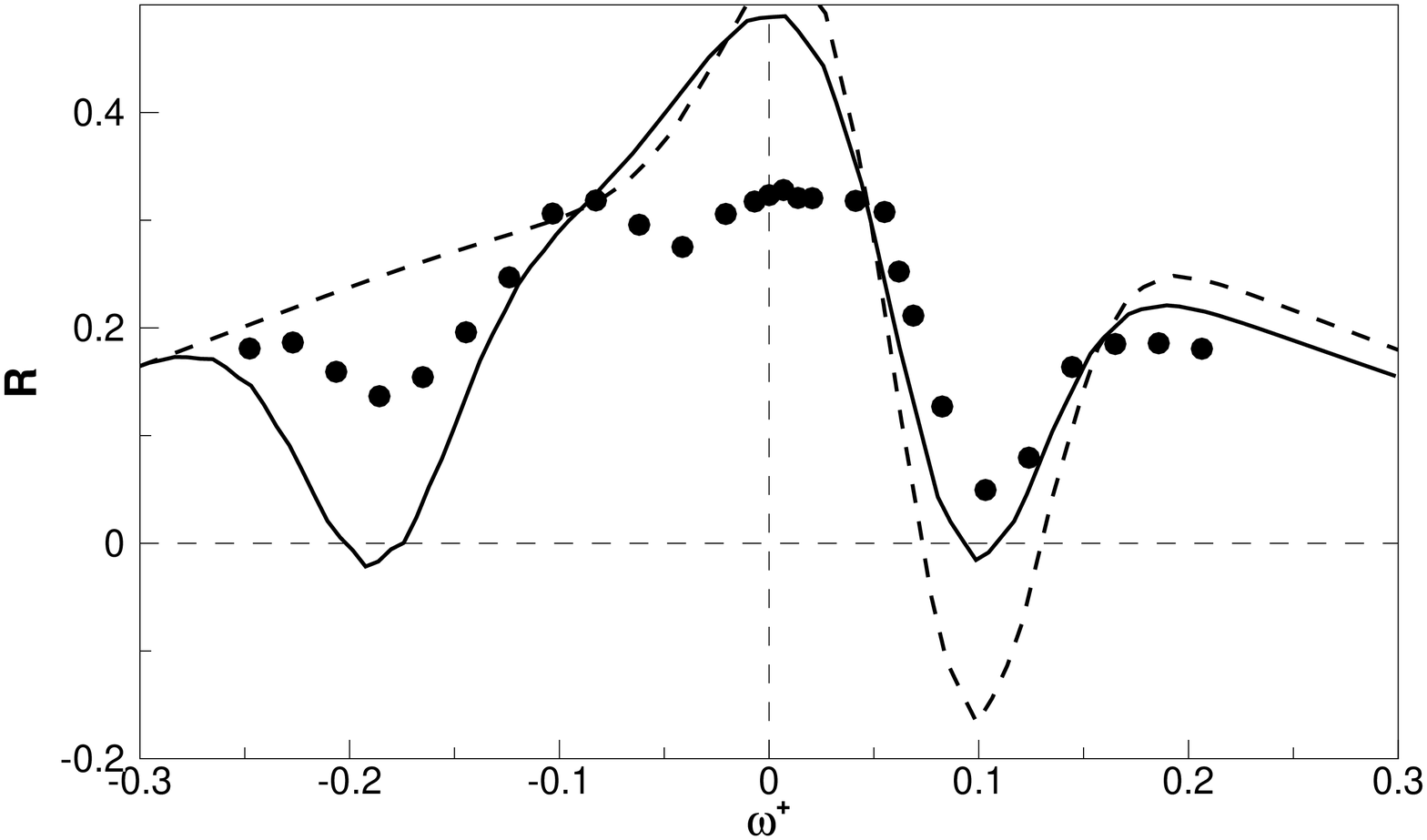}
\caption{Cut of $\tilde{\D}_\ell(\omega,\kappa;3)$ of Fig. \ref{fig:kwmap-p3l} at $\kappa^+=0.0082$, corresponding to $s=3$ (continuous line), compared to experimental data (symbols) and to the DNS data (dashed line).}
\label{fig:p3-l}
\end{figure}

The linear superposition of harmonics again changes the overall shape of the surface describing drag reduction; now $\tilde{\D}$ presents additional patterns and region of local maxima or local minima. Its general shape is shown in Fig. \ref{fig:kwmap-p3l}, where the most evident difference with the function $\D$ is that a region of drag increase is created for backward-traveling waves too. 
An overview of the effects brought about by Eq. (\ref{eq:D-3-1}) can be obtained from figure \ref{fig:wkdiff}, that plots the function $\D(\omega,\kappa) - \tilde{\D}_\ell(\omega,\kappa;3) $ in the entire $\omega-\kappa$ plane. Evident are the prediction of a much smaller drag increase at positive frequencies, as well as a significant decrease of drag reduction in the left half-plane. The dominant changes are identified at the forward speed of about $c^+ \approx 11$ and at the backward speed of twice this value.

Fig. \ref{fig:p3-l} highlights how two key features of the experimental data are reproduced well when the first harmonic is considered. The additional undulation that is present in the experimental data for $\omega^+ \cong -0.18$ now appears in the fitted curve at exactly the same frequency, although the amplitude of the wiggle is different. (Of course, the present linear analysis cannot be expected to predict amplitudes correctly.) Moreover, in the positive frequency branch, adding the effect of the first harmonic significantly reduces the drag increase phenomenon, and helps explaining why an increase in friction drag is not observed in the experiments, that reveal only a sharp drop of drag reduction, although at the correct frequency.

The procedure described so far can be carried on further. Since there is no {\em a priori} reason to assume a linear dependence of drag reduction on the wave components, Eq. (\ref{eq:spectral-superposition}) can be used not only to probe for linear superposition of harmonics, but also to test for non-linear effects. We can for example write that: 
\begin{eqnarray}
\label{eq:nonlin}
&&\tilde{\D}_{n\ell} (\omega,\kappa;3) = C_3 \int\!\!\int \K(\tau,\xi) \left[ f_{\omega,\kappa} + \frac{1}{2} f_{\omega,-2 \kappa} + \right. \nonumber \\
&& \left. + \eta \left( f_{\omega,\kappa} + \frac{1}{2} f_{\omega,-2\kappa} \right)^2 + \ldots \right] \, \ud \tau \ud \xi,
\end{eqnarray}
where the subscript $n\ell$ refers to nonlinear effects, suitably weighted by a parameter $\eta < 1$. The value of $\eta$ is chosen such that the difference between the experimental data and the curve obtained by cutting $\tilde{\D}_{n\ell}(\omega,\kappa;3)$ at the required $\kappa$ is minimized.

\begin{figure}
\includegraphics[width=\columnwidth]{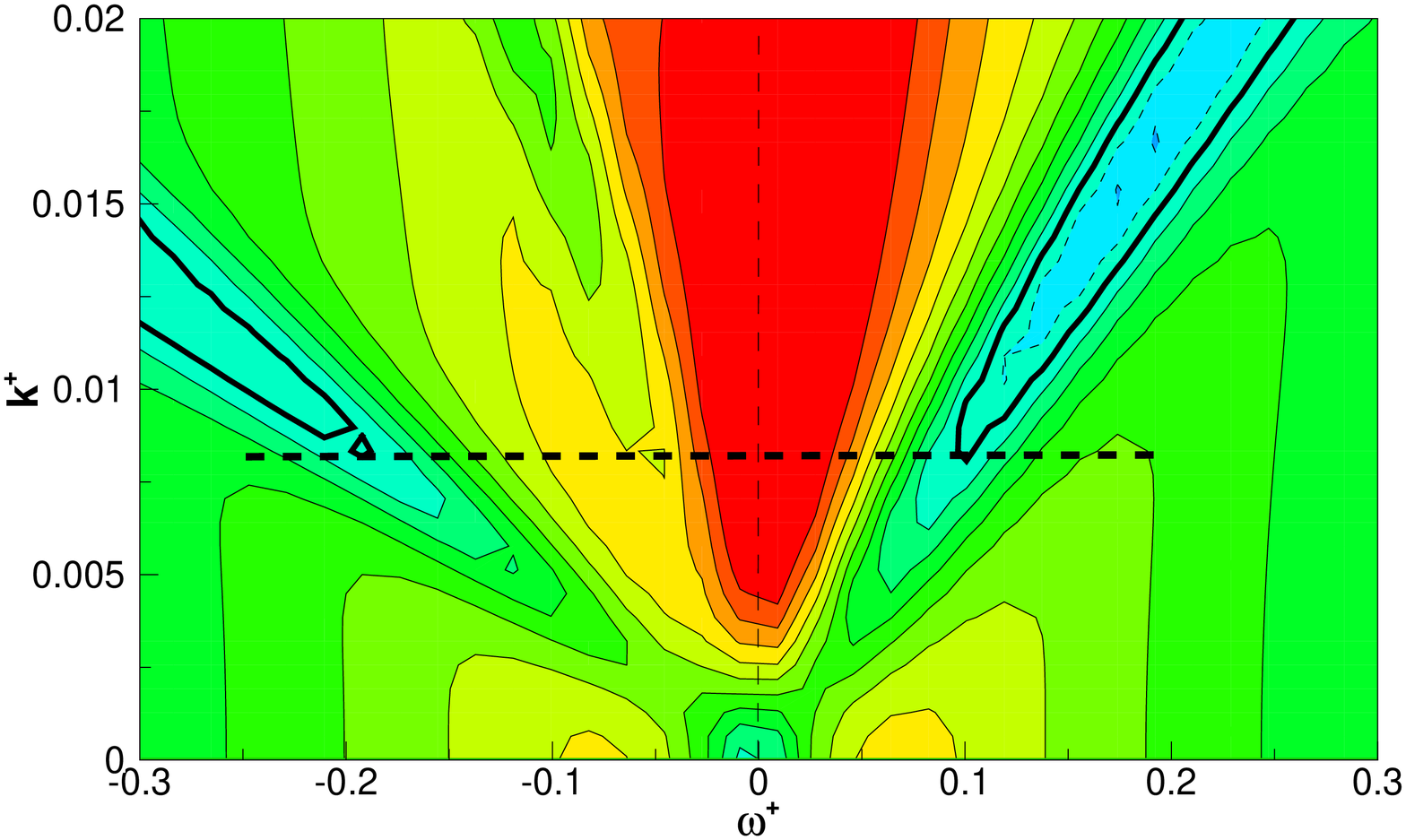}
\caption{Map of the drag reduction rate $\tilde{\D}_{n\ell}(\omega,\kappa;3)$ obtained by formula (\ref{eq:nonlin}), i.e. by adding non-linear effects at first order of the first harmonic, Eq. (\ref{eq:fp3}). Contours as in figure \ref{fig:DNSresults}.}
\label{fig:kwmap-p3n}
\end{figure}

\begin{figure}
\includegraphics[width=\columnwidth]{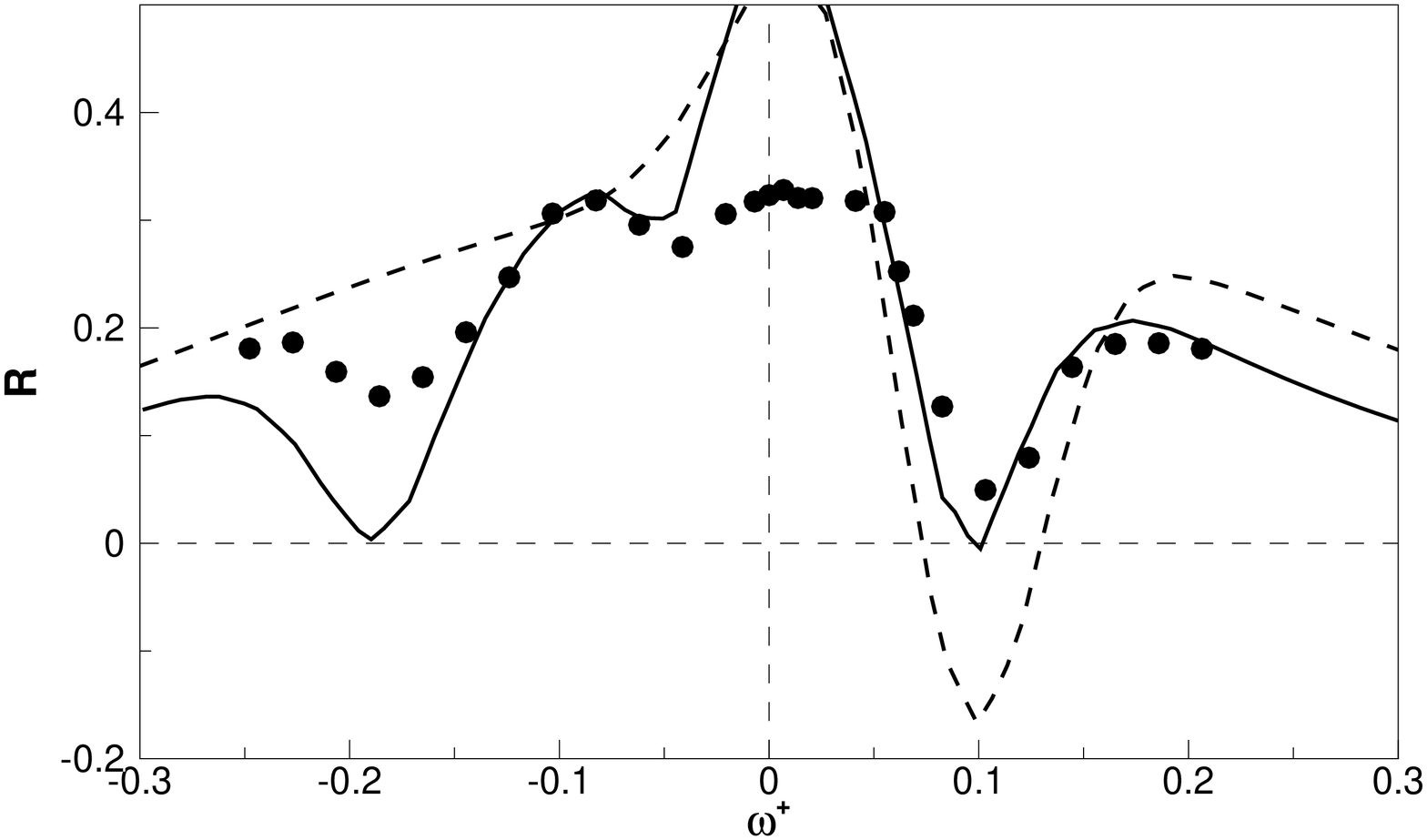}
\caption{Cut of $\tilde{\D}_{n\ell}(\omega,\kappa;3)$ of Fig. \ref{fig:kwmap-p3n} at $\kappa^+=0.0082$, corresponding to $s=3$ (continuous line), compared to experimental data (symbols) and to the DNS data (dashed line). Contours as in figure \ref{fig:DNSresults}.}
\label{fig:p3-n}
\end{figure}

By truncating Eq. (\ref{eq:nonlin}) at second order, i.e. only terms which are first order in $\eta$ are retained, and considering the case $s=3$, the resulting drag reduction map is shown in Fig. \ref{fig:kwmap-p3n}. Its general appearance is rather similar to the previous Fig. \ref{fig:kwmap-p3l}, that was obtained in the linear setting, but a new local drop of drag reduction can be noticed for slow backward-traveling waves. This is better evidenced by a cut of the map for the experimental wavelength corresponding to $s=3$, as shown in Fig. \ref{fig:p3-n}. The nonlinear fit clearly possesses the wiggle of the experimental data for negative frequencies, with a local maximum at $\omega^+=-0.08$ and a local minimum at $\omega^+=-0.04$. 

\begin{figure}
\includegraphics[width=\columnwidth]{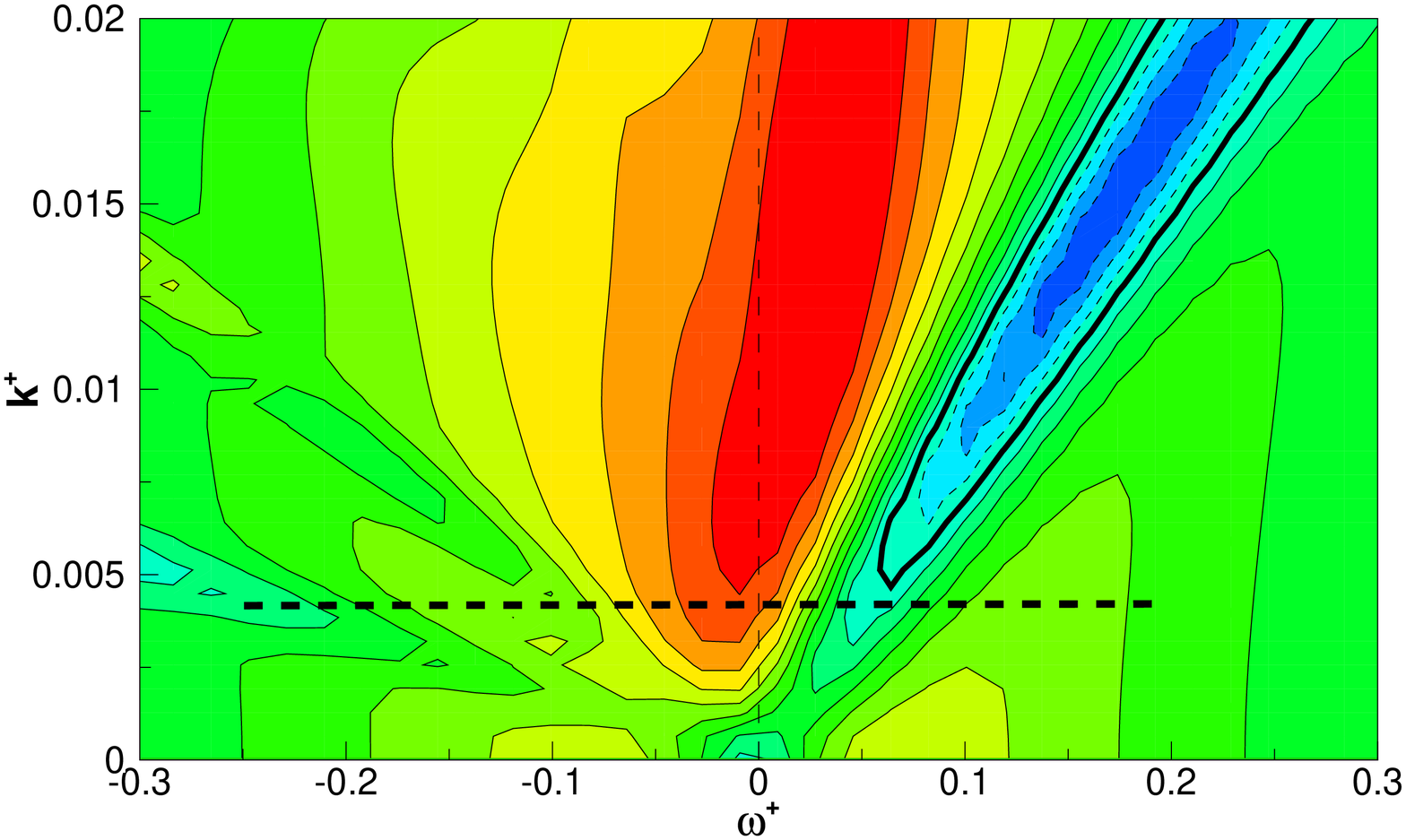}
\caption{Map of drag reduction rate $\tilde{\D}_{n\ell}(\omega,\kappa;6)$ obtained by formula (\ref{eq:nonlin}), i.e. by adding non-linear effects at first order of the first harmonic, Eq. (\ref{eq:fp6}). Contours as in figure \ref{fig:DNSresults}.}
\label{fig:kwmap-p6n}
\end{figure}

\begin{figure}
\includegraphics[width=\columnwidth]{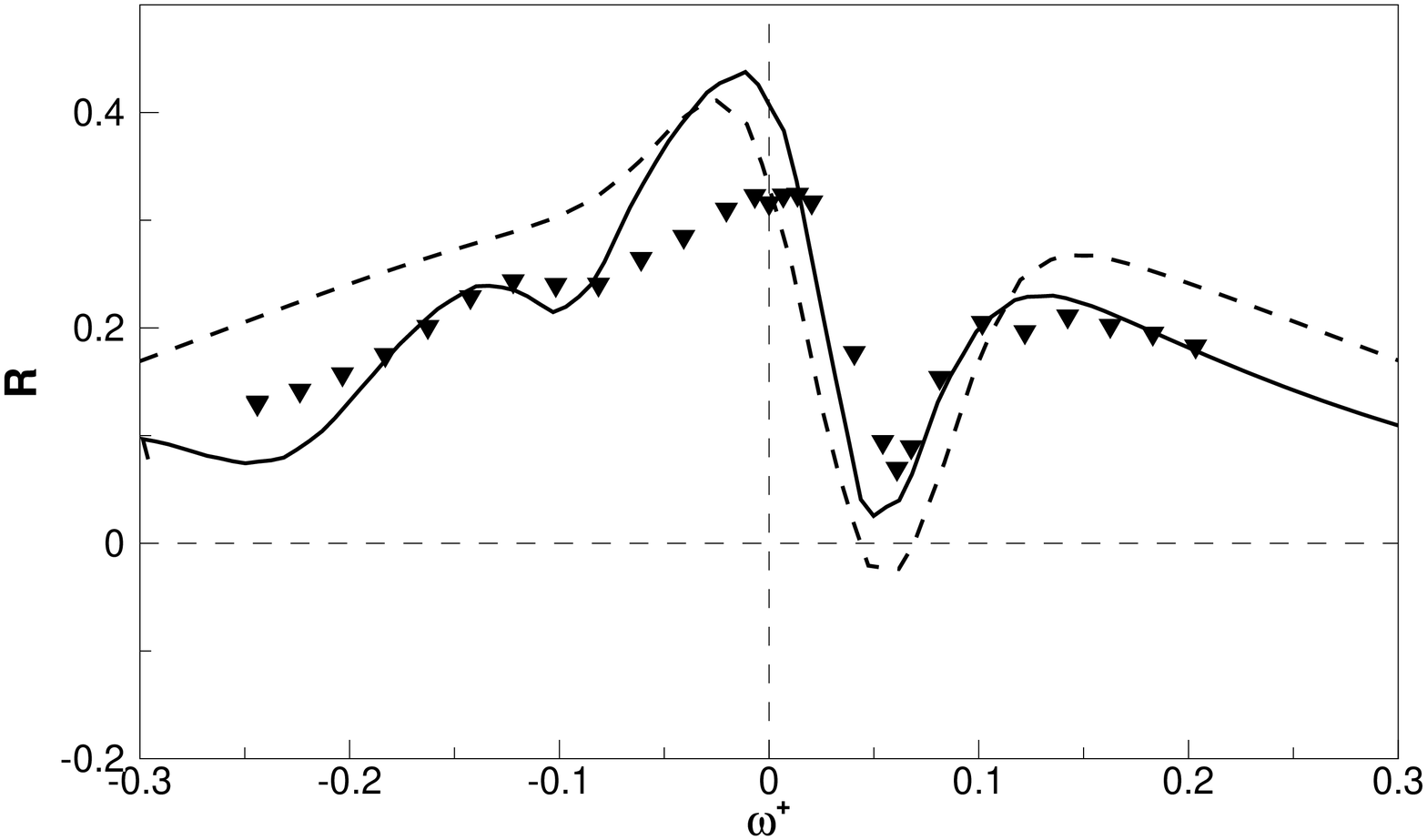}
\caption{Cut of $\tilde{\D}_{n\ell}(\omega,\kappa;6)$ of Fig. \ref{fig:kwmap-p6n} at $\kappa^+=0.0041$, corresponding to $s=6$ (continuous line), compared to experimental data (symbols) and to the DNS data (dashed line).}
\label{fig:p6-n}
\end{figure}

So far we have discussed cases at $s=2$ and $s=3$, for which the discretization of the sinusoidal wave is coarser. By using (\ref{eq:spectral-superposition}) and (\ref{eq:fp6}) the case for $s=6$ can be addressed in a similar way. Plotting the function $\tilde{\D}_\ell(\omega,\kappa;6)$ and its horizontal cut (not shown) reveals similar features. Equation (\ref{eq:nonlin}) for the non-linear case becomes:
\begin{eqnarray}
\label{eq:D-6-1}
&&\tilde{\D}_{n\ell} (\omega,\kappa;6) = C_6 \int\!\!\int \K(\tau,\xi) \left[ f_{\omega,\kappa} + \frac{1}{5} f_{\omega,-5 \kappa} + \right. \nonumber \\
&& \left. + \eta \left( f_{\omega,\kappa} + \frac{1}{5} f_{\omega,-5\kappa} \right)^2 + \ldots \right] \, \ud \tau \ud \xi,
\end{eqnarray}

We only show in Fig. \ref{fig:kwmap-p6n} the plot of the function $\tilde{\D}_{n\ell} (\omega,\kappa;6)$, computed with the same value of $\eta$ previously determined for $S=3$; Fig. \ref{fig:p6-n} shows the horizontal cut of the map together with the experimental data. The general observations discussed above apply here too. 

As a concluding comment, we underline that the present analysis, while essential for interpreting the wiggles in the experimental data as due to discretization effects, does not build on a firm theoretical ground, and it is not capable of an accurate prediction in quantitative terms. This is of course expected, given the simplifying assumptions mentioned above. Fully accounting for quantitative effects would require a full DNS study that is presently beyond our possibilities. However, the present results are useful for designing future, more realistic application-oriented experiments, where correctly accounting for discretization effects will be crucial.

\subsection{Practical issues}
\label{sec:practical-issues}

The present experimental setup, and in particular the strategy chosen to implement the traveling waves, are obviously unsuited for practical applications. However, while briefly addressing issues that are relevant in such practical applications, it is instructive to describe the energetic budget of this active technique, by examining power estimates for two cases of an ideal actuator and of the mechanical actuator used in the present work. 

In our setup, with control off, losses across the entire loop, including those due to the flow meter, require a pumping power of 108 mW, whereas friction losses taking place across the active portion of the pipe are only 6.9 mW. Such small numbers are due to the extremely small value of the bulk velocity. If the waves yielded the maximum $R$ observed in DNS, and were generated by an ideal actuator with unit efficiency, they would be sustained by a power $P_{in}$ of 2.1 mW and would decrease friction by 3.3 mW, thus yielding a net saving with $S=0.17$. In such operating condition, the gain $G$ would be $G=1.6$. If the setup were operated at half the forcing intensity in order to obtain better efficiency, at $A^+=6$ one could ideally obtain 2.4 mW of reduced power with $P_{in}=0.5$ mW, i.e. $S=0.28$ and $G=4.8$. In comparison, the oscillating wall, i.e. a similar active technique that works on a similar physical principle to decrease drag, driven with $A^+=12$ and $T^+=100$ would yield $R=0.33$ with $S=-0.47$ (negative net saving, i.e. net loss).

In practice, figures are very different. Our pump has a rated power of 0.5 kW; it has been chosen without energetic efficiency in mind, and in our setup it has to work extremely far from its design point, that sits at much higher flow rates. Our estimate of the power required to run the experiment is 1,528 mW, which means that the pump wastes most of the energy only to win its internal friction losses. With control on, this power should be reduced to 1,496 mW. Concerning the actuators, the six motors require about $P_{in}=93,000$ mW to run. This translates into negative $S=-13,000$ (i.e. huge net losses), and extremely low gain $G=2.4 \times 10^{-5}$. 

Another issue of practical importance is the behaviour of the traveling waves at higher values of $Re$. There is a lack of reliable data for this and related active drag reduction techniques at values of the Reynolds number high enough to become representative of applications. From a conceptual viewpoint, the very reasonable assumption of near-wall scaling of the phenomenon still needs to be confirmed. At least for the oscillating wall, where some experimental analyses are already available, this could prove delicate, since a slow decreasing trend of $R$ with $Re$ should be detected, and the available data, including DNS, cover only one decade in $Re$. Driving the present setup at higher $Re$ is problematic: the axial size of the moving slabls is obviously fixed in physical units, so that there is no simple way of implementing the orders-of-magnitude larger $Re$ that one would need to estabilish optimal conditions and performance as a function of $Re$.

The last practical aspect that we would like to stress here is the large-scale character of this type of forcing, which implies relatively large spatial and temporal physical scales of actuation even at high $Re$. Following the example \cite{gadelhak-2000} of a commercial aircraft in cruising flight, the spatial scales involved in the traveling waves imply a wavelength of the order of a few millimeters, whereas typical frequencies translate into a few kHz.

\section{Summary and conclusions}
\label{sec:conclusions}

In this paper the large turbulent drag reduction rates that can be obtained by creating streamwise-traveling waves of transverse velocity at the wall of a turbulent duct flow have been experimentally reproduced and confirmed. The 33\% of maximum drag reduction measured here should be regarded as a large and interesting value, that is obtained via a simple large-scale modification of the flow boundary condition, without the need for flow sensors.

The results are in good overall agreement with the sole numerical study available to date. We have also identified several quantitative differences, and discussed the reasons that can explain them. Among several, two important aspects are the different geometry between the numerical study and the present experiment (plane channel flow for the DNS; and circular pipe flow here), and the presence of a spatial transient in the experiment that unavoidably affects a global friction drag measurement as that based on pressure drop. Moreover, an essential aspect that differentiates the present experiment from the DNS simulations is the discrete spatial waveform. A Fourier analysis of the discrete waveform has been carried out to identify the first relevant harmonics that are created when a piecewise-constant velocity distribution is used instead of the idealized sinusoid. Properly accounting for these harmonics, whose effect turns out to be significant, has explained several artifacts of the experimental results that at first sight could be erroneously interpreted as experimental scatter or inaccuracy.

The technique that we have devised here to enforce the traveling wave is obviously suited to a proof-of-principle experiment only. This has been made evident through our discussion of practical aspects of this technique, and in particular by the estimates of its energetic budget. We hope that further technological developments could make such waves worth of pursuing for practical applications, thanks to their attractive intrinsic energetic efficiency. Regardless of this aspect, the streamwise-traveling waves hold great promise for improving our understanding of wall turbulence by shedding new light on the mechanism by which it is sensitive to spanwise forcing. 

\section*{Acknowledgments}

Mr A. Bertolucci and ing. D. Grassi are thanked for their help during the experimental work. Partial financial support from project PRIN 2005 is acknowledged.

\bibliographystyle{unsrt}
\bibliography{../../mq}

\end{document}